\documentstyle[epsfig,referee]{mn}

\newif\ifAMStwofonts
\def\apjs{ApJS}
\def\apj{ApJ}
\def\mnras{MNRAS}
\def\aj{AJ}
\def\aap{A\&A}
\def\aaps{A\&AS}
\def\apss{Ap\&SS}
\def\apjl{ApJL}
\def\pasp{PASP}
\ifoldfss
  \ifCUPmtlplainloaded \else
    \NewTextAlphabet{textbfit} {cmbxti10} {}
    \NewTextAlphabet{textbfss} {cmssbx10} {}
    \NewMathAlphabet{mathbfit} {cmbxti10} {} % for math mode
    \NewMathAlphabet{mathbfss} {cmssbx10} {} %  "   "    "
  \fi
  \ifAMStwofonts
    \ifCUPmtlplainloaded \else
      \NewSymbolFont{upmath} {eurm10}
      \NewSymbolFont{AMSa} {msam10}
      \NewMathSymbol{\upi}     {0}{upmath}{19}
      \NewMathSymbol{\umu}     {0}{upmath}{16}
      \NewMathSymbol{\upartial}{0}{upmath}{40}
      \NewMathSymbol{\leqslant}{3}{AMSa}{36}
      \NewMathSymbol{\geqslant}{3}{AMSa}{3E}

      \let\leq=\leqslant \let\leq=\leqslant
       
    \fi
  \fi
\fi % End of OFSS

\ifnfssone
  \newmathalphabet{\mathit}
  \addtoversion{normal}{\mathit}{cmr}{m}{it}
  \addtoversion{bold}{\mathit}{cmr}{bx}{it}
  \newmathalphabet{\mathbfit} % math mode version of \textbfit{..}
  \addtoversion{normal}{\mathbfit}{cmr}{bx}{it}
  \addtoversion{bold}{\mathbfit}{cmr}{bx}{it}
  \newmathalphabet{\mathbfss} % math mode version of \textbfss{..}
  \addtoversion{normal}{\mathbfss}{cmss}{bx}{n}
  \addtoversion{bold}{\mathbfss}{cmss}{bx}{n}
  \ifAMStwofonts
    \ifCUPmtlplainloaded \else
      \UseAMStwoboldmath
      \makeatletter
      \new@mathgroup\upmath@group
      \define@mathgroup\mv@normal\upmath@group{eur}{m}{n}
      \define@mathgroup\mv@bold\upmath@group{eur}{b}{n}
      \edef\UPM{\hexnumber\upmath@group}
      \new@mathgroup\amsa@group
      \define@mathgroup\mv@normal\amsa@group{msa}{m}{n}
      \define@mathgroup\mv@bold\amsa@group{msa}{m}{n}
      \edef\AMSa{\hexnumber\amsa@group}
      \makeatother
      \mathchardef\upi="0\UPM19
      \mathchardef\umu="0\UPM16
      \mathchardef\upartial="0\UPM40
      \mathchardef\leqslant="3\AMSa36
      \mathchardef\geqslant="3\AMSa3E

      \let\leq=\leqslant \let\leq=\leqslant

    \fi
  \fi
\fi % End of NFSS release 1

\ifnfsstwo
  \DeclareMathAlphabet{\mathbfit}{OT1}{cmr}{bx}{it}
  \SetMathAlphabet\mathbfit{bold}{OT1}{cmr}{bx}{it}
  \DeclareMathAlphabet{\mathbfss}{OT1}{cmss}{bx}{n}
  \SetMathAlphabet\mathbfss{bold}{OT1}{cmss}{bx}{n}
  \ifAMStwofonts
    \ifCUPmtlplainloaded \else
      \DeclareSymbolFont{UPM}{U}{eur}{m}{n}
      \SetSymbolFont{UPM}{bold}{U}{eur}{b}{n}
      \DeclareSymbolFont{AMSa}{U}{msa}{m}{n}
      \DeclareMathSymbol{\upi}{0}{UPM}{"19}
      \DeclareMathSymbol{\umu}{0}{UPM}{"16}
      \DeclareMathSymbol{\upartial}{0}{UPM}{"40}
      \DeclareMathSymbol{\leqslant}{3}{AMSa}{"36}
      \DeclareMathSymbol{\geqslant}{3}{AMSa}{"3E}

      \let\leq=\leqslant \let\leq=\leqslant

      \let\leq=\leqslant \let\leq=\leqslant

    \fi
  \fi
\fi % End of NFSS release 2

\ifCUPmtlplainloaded \else
  \ifAMStwofonts \else % If no AMS fonts
    \def\upi{\pi}
    \def\umu{\mu}
    \def\upartial{\partial}
  \fi
\fi

\title{A comprehensive study of reported high metallicity giant HII regions. II. 
Ionising stellar populations. }
\author[M. Castellanos et al.]
       {Marcelo~Castellanos,$^1$ Angeles I.~D{\'\i}az,$^1$ and Elena~Terlevich$^2$ 
     \thanks{Visiting Fellow, IoA, Cambridge}\\
        $^1$ Departamento de F{\'\i}sica Te{\'o}rica, C-XI, Universidad Aut{\'o}noma
        de Madrid, 28049 Madrid, Spain\\
        $^2$INAOE, Tonantzintla, Apdo. Postal 51, 72000 Puebla,
        M{\'e}xico\\}

\pagerange{\pageref{firstpage}--\pageref{lastpage}}
\pubyear{2000}

\begin{document}

\maketitle

\label{firstpage}

\begin{abstract}

  The ionising stellar populations  of eleven H{\sevensize II} regions
in the spiral galaxies: NGC~628, NGC~925, NGC~1232 and
NGC~1637,  all 
of them reported to have solar or oversolar abundances according to
empirical calibrations, have been analysed using stellar population
synthesis models. 

Four of the observed regions in the sample, show features which
indicate the presence of a population of Wolf-Rayet 
(WR) stars with ages between 2.3 and 4.1 Myr. 
This population is sufficient to explain the emission line spectrum of
the low metallicity region H13 in NGC~628, taking into account the
uncertainties involved in both observations and model computations. 
This is not the case for the rest of the regions for which
a second ionising population is required to simultaneously reproduce
both the WR
features and the emission line spectrum. 
Composite populations are also found for half of the regions without WR
features, although in this case, the result is based only on the
emission line spectrum analysis.

For two of the regions showing WR features,
no consistent solution is found, since the
population containing WR stars produces a spectral energy distribution
which is too hard to explain the emission of the gas. Several solutions
are proposed to solve this problem.
\end{abstract}


\section{Introduction}

Metallicity affects stellar evolution in at least two different ways:
by increasing the opacity of the stellar material and through the
strengthening of the wind driven mass loss in high mass stars. The
former implies a lower effective temperature for the atmospheres of
ionising stars of higher metal content even at zero age, while the
latter can severely affect the evolution of the most massive stars
leading, in the most extreme cases, to the almost complete loss of
their outer envelopes.

The first of these effects should be readily observable as an inverse
correlation between metallicity and temperature of the ionising
radiation for HII regions of different metallicities. In fact, there is
a general agreement about the hardening of the ionising radiation in
regions of low metal content (e.g. Campbell et al. 1986; Cervi{\~n}o and
Mas-Hesse 1994). The inverse situation, however, i.e. the softening of
the ionising radiation in regions of high metal content, is more
difficult to establish since, in these cases, the determination of both
metal abundance and ionising temperature is rather difficult. For the
first one we have to rely , in general, on uncertain semi-empirical
calibrations when the involved abundances are higher than about 1/2
solar. For the second one, there are no suitable nebular emission line
diagnostics. Recently, Bressolin, Kennicut \& Garnett (1998) have
investigated a possible relation between ionising stellar temperature
and metallicity in giant HII regions in spiral galaxies through the use
of the softness parameter $\eta$' (V{\'\i}lchez \& Pagel 1988) concluding that
their results are consistent with a significant decrease in mean
stellar temperatures of the ionising stars with increasing
metallicity. This decrease is however difficult to quantify since in
this work the metallicity is derived from the abundance parameter
R$_{23}$ (Pagel et al. 1979) and both  $\eta$' and R$_{23}$ depend to some
extent on the degree of ionisation of the nebula (D{\'\i}az et al. 1991). On
the other hand, the method can be applied to a large number of HII
regions and therefore can provide trends statistically significant.

A different approach consists in deriving the metal abundance from
the direct determination of the electron temperature in the nebula and
the ionising temperature from the fitting of single star
photoionisation models. The method requires both high quality
spectroscopy and detailed modelling. In a previous paper (D{\'\i}az et
al. 2000a; DCTG00) we have applied this procedure to HII regions in the spiral
galaxy NGC~4258. The combination of these data with similar ones in the
literature also seems to point to a decrease of mean ionising
temperature with increasing metallicity. Surprisingly, this is so even
in the presence of WR stars.

The presence of these stars provides an additional constraint to
characterise the ionising stellar population. In the last years,
stellar evolution models for massive stars taking into account mass
loss at different metallicities have been calculated in order to
reproduce the observed features of WR stars. The model predictions are
however quite different depending on the assumed mass loss. In fact,
models that assume a standard mass loss rate underpredict the number of
galactic WR stars in comparison with observations (Maeder \& Meynet
1994) while an enhanced mass loss rate fits most WR population
properties except for the mass loss rate itself (Leitherer, Chapman \& 
Koribalski 1997). Therefore, the detailed observation and modelling of
giant HII regions showing WR features can help to shed some light on
this very important issue.
 
It would also be desirable to check the hardening of the ionising
radiation due 
to the presence of Wolf-Rayet stars predicted on theoretical grounds
(P{\'e}rez 1997). These stars are supposed
to change drastically the spectral energy distribution of an ionising
star cluster at energies higher than 
{$\sim$} 40 eV. Models predict that this change depends both on 
the age and the metallicity of the burst (Schaerer \& Vacca 1998; SV98) 
and, as a result of this, ionic elemental ratios are expected to change as 
well as ion-weighted temperatures.

 A puzzling result in DCTG00
was the relatively low effective temperatures found in the two
ionising clusters with WR features, suggesting either that the rate of
high energy photons is lower than expected or that the opacity in WR
envelopes is highly efficient. 
 This 
matter can be investigated through the observation of WR features in high 
metallicity HII regions where the effect of the mass loss rate on the 
effective temperature of the ionising stars is supposed to be most enhanced. 
 
In order to analyse the effects of metallicity on both  stellar evolution
and ionising radiation, we have analysed a sample of reported high 
metallicity H{\sevensize II} regions by combining evolutionary
synthesis models for ionising populations and  photoionisation models
in a selfconsistent way.  In this paper a detailed analysis of the
ionising populations of the observed regions is presented.

\section{Physical properties of the observed HII regions}

The observations of the sample HII regions analysed here: 4 in NGC~628,
2 in NGC~925, 4 in NGC~1232 and 1 in NGC~1637  are described in
detail in Castellanos, D{\'\i}az \& Terlevich (2002; Paper I) and consist of 
spectrophotometry
in the 4000-9700 {\AA} range. These data were complemented with those by
Van Zee et al. (1998) in order to provide a complete set of emission
line ratios in the spectral range from 3500 to 9700 {\AA}. For every region 
the physical conditions of the ionised gas-- particle density, electron
temperatures, ionic and total abundances -- and functional parameters
-- ionisation parameter and temperature of the ionising radiation -- as
derived from the application of single star photoionisation models were
determined and are summarized in Table 1 for the regions analysed here.

%%%%%%%%%%% H alpha fluxes and luminosities

For each of them we have calculated the H$\alpha$
luminosity, L(H$\alpha$), the number of hydrogen ionising photons, 
Q(H), the angular effective diameter, $\phi$, the filling
factor, $\epsilon$, the 
mass of ionised hydrogen, M(HII), and the mass content in the ionising 
clusters, M$^{*}$. 
These quantities have been derived according to the expressions given
in D{\'\i}az (1998), for our adopted distances: 7.3 Mpc for NGC~628, 
8.6 Mpc for NGC~925, 21.5 Mpc for NGC~1232 and 7.8 Mpc for NGC~1637
(see Paper I) and are listed in Table 2. 

The H$\alpha$ fluxes for regions H13, H3, and H5, uncorrected for reddening,
are 669, 95 and 62 $\times$ 10$^{-16}$ erg s$^{-1}$ cm$^{-2}$ respectively,
a factor of about three lower than those measured by Kennicutt \& Hodge (1980)
from H$\alpha$
photometry (2089, 372 and 145 $\times$ 10$^{-16}$ erg s$^{-1}$
cm$^{-2}$). This factor reduces to about 2.4 if we take into account
that their H$\alpha$ filter also includes the [NII] lines which amount to
about 1/3 of H$\alpha$. 

%%%%%%%%%%%%%%%%% Comments in other sources of photometry

H$\alpha$ photographs (Hodge 1976) reveal these regions to be
rather extended, and considering that our observations have been obtained
through a 1\farcs03 width slit, this comparison implies that our
derived values should be considered as lower limits. At any rate, our
value for H13 agrees well with that measured by Van Zee et al. (Van
Zee, private communication) which was obtained through a 2 \farcs0
diameter aperture.

For the rest of the regions we have not found any published photometric
data.

%%%%%%% Sizes 

The angular diameter of the emitting regions as well as the filling factors
can be obtained from 
the ionisation parameter, u, the reddening corrected H$\alpha$ flux, F(H$\alpha$)
and the derived electron density, n$_H$, under the assumption of
spherical geometry. Then 

\[ u = Q(H)/4\pi c R^2 n_H \]

\noindent
where Q(H) is the number of hydrogen ionising photons per
second, R is the radius of the ionised region and c is the speed of
light.

The corresponding angular diameter is then found to be:
$$ \phi  =
0.64\left(\frac{F(H\alpha)}{10^{-14}}\right)^{1/2}\left(\frac{u}{10^{-3}}\right)^{-1/2}\left(\frac{n_H}
{10^{2}}\right)^{-1/2}$$

\noindent
where F(H$\alpha$) is in units of erg cm$^{-2}$s$^{-1}$, n$_H$ is in units of
cm$^{-3}$ and $\phi$ is in arcsec. For the regions for which only upper limits to the
electron density could be derived, a value of n$_H$=40 cm$^{-3}$ has
been assumed. Angular diameters thus calculated, do not depend on
the assumed distance to the galaxy. For all the regions, we obtain angular
diameters between 0.6 and 2 arcsec.

%%%%%%%% Filling factors

Filling factors for each observed region can also be determined from
the reddening corrected  H$\alpha$ flux and the derived ionisation parameter, 
according to the expresion:
$$ \epsilon =
0.19\left(\frac{u}{10^{-3}}\right)^{3/2}\left[\left(\frac{F(H\alpha)}{10^{-14}}\right)\left(\frac{n_H}{10^{2}}\right)\right]^{-1/2}
\left[\left(\frac{\alpha_B(H^{\rm o},T)}{10^{-13}}\right)\left(\frac{D}{10}\right)\right]^{-1}$$

\noindent
where D is in Mpc and $\alpha_B(H^{\rm o},T)$ is the case B recombination
coefficient for hydrogen for which we have used a value of 3.76 $\times$
10$^{-13}$ cm$^{3}$ s$^{-1}$, corresponding to T= 7000 K and n$_H$= 100 cm$^{-3}$ 
(Osterbrock 1989). Filling factors thus derived depend inversely on the
galaxy distance, D. 
Values between 0.008 and 0.13 are found for our adopted distances. 
These values are very similar to those derived for the observed HII
regions in NGC 4258 (see DCTG00) and M~51 (D{\'\i}az et al. 1991).

%%%%%% Ionised hydrogen masses

The corresponding masses of ionised hydrogen range from 1.5 $\times$
10$^{4}$ M$_{\odot}$ for region H4 in NGC~628 to 8.8 $\times$ 10$^{4}$ M$_{\odot}$ for 
region CDT4 in NGC~1232.

%%%%%% Masses of the ionising stellar clusters

Regions CDT1, CDT3 and CDT4 in NGC 1232 have H$\alpha$ luminosities greater 
than 10$^{39}$ erg s$^{-1}$ and hence 
can be classified as supergiant HII regions as defined by Kennicutt (1983). 
The rest of the regions have H$\alpha$ luminosities typical of HII regions 
in early spiral galaxies, although all of them are greater than 10$^{37}$ 
erg s$^{-1}$ requiring more than a single
star for their ionisation (Panagia 1973).

For our observed regions, all values for Q(H) are greater than
10$^{49}$ photon s$^{-1}$. 
For regions  CDT1, CDT3 and CDT4 in NGC~1232  
values as high as 10$^{51}$ photon s$^{-1}$ are found. In the absence of
dust and if no photons are scaping the region, a lower limit for the
mass of the ionising clusters can be 
estimated by means of the H$\beta$ measured equivalent width and the H$\alpha$
luminosity for each region (see D{\'\i}az 1998). Assuming a Salpeter IMF
with upper and
lower mass limits of 100 and 0.8 M$_{\odot}$ respectively, the estimated
ionising cluster masses range
from 1.7 $\times$ 10$^{3}$ M$_{\odot}$ for region H4 in NGC 628 to 1.80 $\times$
10$^{5}$ M$_{\odot}$ for region CDT1 in NGC~1232. It can be inferred from 
these results that all the regions are ionised by relatively small
clusters, except for the three supergiant HII regions in NGC
1232. Regions H3, H4 and
H5 show a ten to one ratio between their ionised hydrogen mass and the mass 
of the ionising stellar clusters which may be then very young. 
Their high observed EW(H$\beta$) values of about  200 {\AA} support this claim. 

If a given fraction of ionising photons were absorbed
by dust grains before they reach the neutral
hydrogen, then the number of Lyman continuum photons emitted Q(H)$_{em}$ by the star
cluster would be higher : Q(H) = f .Q(H)$_{em}$ with f $<$
1. Therefore, our assumption of f = 1, {\it i. e} absence of dust,
implies that Q(H)$_{em}$ and correspondingly M* are lower limits. The rest
of the derived nebular parameters: angular diameter, filling factor and
mass of ionised hydrogen depend on the extinction corrected H$\alpha$
flux, the ionization parameter and the hydrogen density of the emitting
gas. All these quantities being observables, they are not affected by
the presence of absorbing dust. If f is less than unity, Q(H)$_{em}$ would be
larger than the derived Q(H) and obviously if all these emitted photons were ionising
the gas we should observe a much higher H$\alpha$ flux and the nebular parameters would be different.

\subsection{Wolf-Rayet populations}

%%%%%%%%%%%%%%%%%%%%%%%%%%%%%%%%%%%%%%%%%%%%%%%%%%%%%%%%%%%%%%%%%%%%%%%%%%%%%
%                         WR en H13 de NGC 628 
%%%%%%%%%%%%%%%%%%%%%%%%%%%%%%%%%%%%%%%%%%%%%%%%%%%%%%%%%%%%%%%%%%%%%%%%%%%%%

Relatively prominent Wolf-Rayet features have been identified in several
of the observed HII regions:  H13 in NGC~628, CDT3 and, at
a fainter level, CDT1 and  CDT4 in NGC~1232. In all of them, the
typical blue `bump' around $\lambda$ 4660 {\AA} is observed and a red `bump'
around $\lambda$ 5800 {\AA} is also present in regions H13 (NGC~628) and CDT3 
(NGC~1232). Details concerning these WR features can be found in
Paper I and are summarized in Table 3.

\subsubsection{Region H13 in NGC~628}
Assuming a distance to NGC~628 of 7.3 Mpc (Sharina et al. 1996), and a
constant extinction value through this region of c(H$\beta$) = 0.29, as
derived from the Balmer and Paschen recombination lines, the total 
luminosities of the blue and red bumps are (1.8 $\pm$ 0.2) $\times$ 10$^{37}$ erg s$^{-1}$
and (6.5 $\pm$ 0.6) $\times$ 10$^{36}$ erg s$^{-1}$ respectively. The former
value comprises the features of N{\sevensize III} 
$\lambda\lambda$ 4634, 4640, and He{\sevensize II} $\lambda$ 4686 {\AA} lines. The
contribution of the N{\sevensize III} lines to the blue WR bump is 
metallicity-dependent according to Smith (1991). In our case this 
contribution represents 0.45 times the total emission. Hence, the
derived value for the stellar HeII line luminosity is: 
\[ L(He{\sevensize II} \lambda 4686) = (9.5 \pm 0.4) \times 10^{36} ergs^{-1} \]
Based on the characteristics of the WR features we can conclude that
the observed WR stars can be classified as WN7 and 
using the calibration of Vacca \& Conti (1992), 6 WN7 stars
are found in region H13.

%%%%%%%%%%%%%%%%%%%%%%%%%%%%%%%%%%%%%%%%%%%%%%%%%%%%%%%%%%%%%%%%%%%%%%%%%%%%%%%%
%%%%%% WR en 1232b2 (CDT3) de NGC 1232 %%%%%%%%%%%%%%%%%%%%%%
%%%%%%%%%%%%%%%%%%%%%%%%%%%%%%%%%%%%%%%%%%%%%%%%%%%%%%%%%%%%%%%%%%%%%%%%%%%%%%%%

\subsubsection{Region CDT3 in NGC~1232}
In this case,  the total luminosities  of the blue and red bumps,
assuming the distance 
to NGC~1232 of 21.5 Mpc adopted by Van Zee et al. (1998) and our
derived 
reddening constant  c(H$\beta$) = 0.32, are (4.3 $\pm$ 1.0) $\times$ 10$^{37}$ erg s$^{-1}$
and (9.4 $\pm$ 2.5) $\times$ 10$^{36}$ erg s$^{-1}$ respectively. 
The derived value for the stellar HeII line luminosity is:
\[ L(He{\sevensize II} \lambda 4686) = (1.95 \pm 0.50) \times 10^{37} ergs^{-1} \]
We classified the dominant WR stars as of WN8 type. Therefore
using the calibration of Vacca \& Conti (1992), 12 WN8 stars
are found in this region. The detection of CII, CIII at $\lambda$ 5140{\AA} could be explained
 by the presence of early or intermediate WC stars. It is only possible to give a lower limit
to the number of WC stars because it is difficult to establish the dominant WC type (SV98). Hence,
we have adopted a mean value for the average observed luminosity of WC4 stars at $\lambda$ 5808{\AA}
of 3.0 $\times$ 10$^{36}$ erg s$^{-1}$. Therefore, at least 3 WC stars could be found in this region. 
\subsubsection{Rest of the regions in NGC~1232}
Fainter Wolf-Rayet features are found in regions CDT1 and CDT4  in 
NGC 1232. Region CDT1 shows a stellar HeII line 
luminosity of (9.5 $\pm$ 1.0) $\times$ 10$^{36}$ erg s$^{-1}$ 
compatible with the presence of 6 WN8
stars.

For region CDT4, the measured HeII line luminosity 
is (5.0 $\pm$ 0.8) $\times$ 10$^{37}$ erg s$^{-1}$. Accordingly,
30 WN8 stars would be necessary to explain the observed luminosity.

\section{Modelling the ionising populations}

The time evolution of the ionising energy output of a young star cluster
can be followed with the help of stellar evolutionary tracks combined
with adequate atmosphere models and they must 
incorporate the most massive stars. This has been made in the last years by
several independent groups (e.g.  Cervi{\~n}o \& Mas-Hesse 1994;
Garc{\'\i}a-Vargas, Bressan \& D{\'\i}az 1995; Leitherer \& Heckman 1995). In these
so called ``evolutionary synthesis models'' the ionising stellar
population is represented by a cluster of coeval stars formed with a
given IMF which evolve in time along to theoretical evolutionary
tracks in the H-R diagram and that is able to ionise the surrounding
gas. Then, the combination of the individual stellar spectral energy
distributions (SED) along a given isochrone representing a given
cluster age is computed. These individual SED can be taken directly
from observations or from stellar atmosphere models. Although the
method is rather complex and the assumption of different stellar
evolution and atmosphere models can lead to different predictions for the SED
of a given cluster, there are several important conclusions which are
common to the work by all groups  mentioned above. In particular, a 
hardening of the
ionising spectrum is predicted to occur between 3 and 4 Myr after the
formation of the cluster as a result of the appearance of WR
stars. In fact, the presence of these stars, which can be detected by
the conspicuous high ionisation emission lines produced in their
powerful winds (see section 2 above) provides important clues to
constrain the age of the ionising population.

SV98 have presented very detailed models of the WR
population in young star clusters, at different metallicities from
Z=0.001 (1/20 Z$_{\odot}$) to Z=0.04 (2 Z$_{\odot}$). Their clusters are formed
according to a Salperter IMF with upper and lower mass limits of 120
M$_{\odot}$ and 0.8 M$_{\odot}$ respectively. They use the stellar
evolution models by Meynet et al. (1994) which assume a mild
overshooting and  enhanced mass loss rate. These models have been shown
to reproduce the observed WR/O star ratios in a variety of regions
(Maeder \& Meynet 1994). Regarding the energy output of main sequence stars
they use the CoStar models (Schaerer \& de Koter 1997) which include 
non-LTE effects, line
blanketing and stellar winds, for stars with initial masses larger than
20 M$_{\odot}$ and Kurucz (1992) plane-paralel LTE models, including line
blanketing effects, for less massive stars. The atmospheres of evolved
stars in the WR phase correspond to the spherically expanding, non-LTE,
unblanketed models by Schmutz, Leitherer \& Gruenwald (1992).   

SV98 provide accurate predictions, as a function of the
cluster age, for the total number of WR stars and their subtype
distribution, the broad stellar emission lines and the luminosities and
equivalent widths of the two ``WR bumps'' at $\lambda$ 4650 and $\lambda$ 5808 {\AA} . We 
have used these models to derive the age of the ionising population which
contains WR stars and whose metallicity has been previously derived. 

Regarding the integrated SED from the ionising cluster, the recent
models from Leitherer et al. (1999; STARBURST99) provide an almost
selfconsistent frame to be used in combination with the WR models
described above. In fact, they use the same stellar evolution models
with enhanced mass loss rate, the same atmosphere models to describe
the stars in the WR phase and cover the range of metallicities and IMF
used by SV98. The only appreciable difference between both
sets of models concerns the atmospheres of the main sequence stars with
initial masses greater than 20 M$_{\odot}$ which in STARBURST99 are
represented by the plane-paralel Kurucz's models implemented by 
Lejeune, Cuisiner \& Buser (1997). We have
therefore used the STARBUST99 models in order to fit the emission line
spectra for our analysed regions.

\subsection{Model fits to the emission line spectra}

% Model description
% Input: O/H, ne, u, T*
% Observables: EW(Hbeta), L(Hbeta), line emission spectra
% Optimisation for u and cluster age
% Results
% Discuss on the basis of Bresolin et al. 1998 and Dopita et al. 2000

%For each of the analysed regions we have the following set of derived
%parameters: H$_\alpha$ luminosity, gas elemental abundances relative to
%hydrogen, electron density, filling factor, ionisation parameter and
%temperature of the ionising radiation according to Mihalas (1972)
%stellar atmospheres (see details in Paper I).

In order to predict the emission line spectrum for each analysed region,
SEDs predicted by the STARBURST99 models for the stellar metallicity
closest to the mean gas abundance were used as input to the latest
version of CLOUDY (Ferland 1999). The models also required the gas
electron density (assumed to be constant through the nebula) and
elemental 
abundances and  the ionisation parameter, all of
which had been previously derived (see Table 1 and Paper I for 
details). In all cases, we have used as input to the photoionisation models
the derived values of the ionisation parameter, electron density and 
epsilon and an inner radius adequate
to reproduce the derived ionising photons. In all the cases the
thickness of the ionised gas shell is less than 10\% of the
total dimensions of the region. This results in a plane-parallel geometry.
 
We have run models with values of the  ionisation parameter inside
$\pm$ 0.5 dex around the derived value and cluster ages with
corresponding equivalent temperatures inside $\pm$ 2000 K around the
ionisation temperature estimated from single star photoionisation
models. 

The equivalent temperature for a given star cluster has been
taken as the stellar effective temperature, from Mihalas (1972) model 
atmospheres, providing the same 
helium-to-hydrogen  ionising photon ratio. In all cases the nebulae are
assumed to be ionisation bounded. Given the run of equivalent
temperature with age for ionising clusters,which shows an important
increase when the most massive stars enter the WR phase, usually more than one
solution is possible. 
This can be seen in Figure 1 where we show, for different metallicities
0.2, 0.4 times solar and solar, the cluster equivalent temperature as a
function of age.  

The best fitting models are then obtained by using an optimisation
method that includes the more intense emission line ratios relative to
H$\beta$. Our models do not explicity treat the physical processes associated 
with the presence of dust grains. However, since  gas-phase measured 
abundances are used as input to the models, the effect of
depletion onto dust grains is taken into account. The
presence of dust grains will affect the thermal equilibrium of the
nebula, by increasing the electron temperature and producing an
enhancement of the optical forbidden lines. Nevertheless, 
according to Shields \& Kennicutt (1995), the depletion of gas
phase coolants is likely to be the most important aspect of dust in
terms of its influence on the observed spectrum.

Tables 4 to 14 give, for each region, the predicted line intensities and
abundances for
these models together with the observed ones and their corresponding
errors. As can be seen most regions show emission line spectra which
can be produced by stellar clusters of ages between 2 and 3 Myr,
although in some cases other solutions are also possible.

\subsection{WR populations}

\subsubsection{Region H13 in NGC~628}

The observed WR blue and red `bump' luminosities 
relative to H$\beta$ are 0.08 and 0.03 respectively. According to the
models by SV98, maximum values are 
found for an age of 4Myr (0.06 and 0.03 for the blue and red `bumps'
respectively). 
Moreover, at the age of 4Myr, the blue and red `bumps' equivalent 
widths are at their maxima, with values around 8.5 {\AA} and 8 {\AA}
respectively. 
These values are in excellent agreement with the observed ones: 8.9 {\AA} for the
blue bump and 
6.1 {\AA} for the red one. Regarding the 
individual lines, the observed values of L(HeII)/L(H$\beta$) and EW(HeII) are
0.04 and 4.8 {\AA} 
respectively. Again, the agreement can be considered 
satisfactory for a single stellar population  of about 4Myr (0.02 and 3 {\AA} 
respectively for an age of 4.1 Myr). Finally,
models predict  H$\beta$ equivalent widths of 173 {\AA} at 4Myr and
159 {\AA} at 4.1 Myr, 
to be compared with the observed value of 140 {\AA}. Taking into account 
the global and individual properties of both `bumps', the agreement 
between the models and the derived quantities is excellent.

Therefore we can assign an age of 4-4.1 Myr to the ionising stellar
population which would be responsible for the WR emission features.

\subsubsection{HII regions in NGC~1232}

From the four HII regions of NGC 1232 analysed in Paper I, three of
them present weak 
blue WR `bumps' (CDT1, CDT3 and CDT4). Special attention must be
paid  to region CDT1 showing a metallicity close to solar. For this
region the observed WR blue `bump' luminosity relative to H$\beta$ is 0.04,
extremely low for a high metallicity HII region. A very low value is
also found for the blue `bump' equivalent width, near 2 {\AA}. Regarding
the stellar HeII luminosity and equivalent width, values of 0.015 and 
0.8 {\AA} are found. According to models by SV98, these
low values could be produced by an evolved ionising population with an
age around 7 Myr. This population would provide a blue ``bump''
equivalent width of
2.4 {\AA}, consistent with the observed value, but would overestimate its
luminosity relative to  H$\beta$ by more than an order of magnitude (0.55 {\it vs}
0.04), while providing an H$\beta$ equivalent width of only 4 {\AA} {\it versus}
the observed one of 48 {\AA}.

 Low values of blue `bump' luminosities and equivalent widths
are also found for early ages, at the beginning of the WR phase. A
cluster of 2.3 Myr of age provides values of the blue `bump' luminosity
0.02 with respect to H$\beta$
and 5.17 {\AA} for its equivalent width, roughly consistent with the observations. This
population, however, would produce an equivalent width of 
H$\beta$ of 312 {\AA}, much higher than observed.   

Similar young ages are found when this sort of analysis
is applied to regions CDT3 and CDT4. Region CDT3 shows a WR `bump' luminosity 
relative to H$\beta$ of 0.03. Taking into account its derived metallicity
(12 + log O/H = 8.56 or Z = 0.008), SV98 predicts this  value for a
cluster with an 
equivalent width of H$\beta$ of {$\sim$} 200{\AA}, which is in excellent 
agreement with our measured value of 189 {\AA}. This corresponds to an
instantaneous burst with age between 3.1 and 3.5 Myr. Furthermore, 
the predicted blue `bump' equivalent width at this age ranges from
5.5 to 10 {\AA} in good agreement with the observed one (5.9 {\AA}). Consistency
is also found for the weak red `bump', whose observed values for both the 
luminosity and equivalent width are 0.007 and 2.0 {\AA} respectively that
should be  
compared with the predicted ones (0.006-0.015 and 2.2-4.2 {\AA}
respectively). Regarding
the HeII line luminosity and equivalent width, our observed values
are 0.015 and 2.7 {\AA} in excellent agreement, again, with the predicted
ones (0.011-0.007 and 2.15-0.75 {\AA} respectively).

In the case of region CDT4, the WR blue `bump' luminosity relative to
H$\beta$ is 0.08.
For its  derived metallicity (12 + logO/H = 8.37 or Z = 0.006). A
value of 0.11 is predicted for an instantaneous burst of 4.1 Myr at
Z=0.008, 
which in
turn predicts an H$\beta$ equivalent width of  105 {\AA}. This value is 
consistent with the observed one of 138 {\AA}. The observed WR blue `bump'
equivalent width is 10 {\AA} in good agreement with the predicted value of
10.8 {\AA} at this age. The HeII line intensity and equivalent width are
again fully consistent
with the corresponding predicted values  (0.03 {\it versus} 0.04 and 3.9
{\AA} {\it versus} 3.7 {\AA}, respectively). 

Therefore relatively young single ionising populations between 3 and 4
Myr are required to
reproduce the observed WR features in regions CDT3 and CDT4. For region
CDT1, no consistent solution is found. An old cluster of 7 Myr would
roughly fit the equivalent width of the WR blue `bump' 
but would overpredict its luminosity with respect
to H$\beta$ by more than  an order of magnitude and underpredict the H$\beta$
equivalent width by about the same amount.  On the other hand, young
clusters of about 2.5
Myr predict  WR blue `bump' luminosities and equivalent widths
in agreement with observations but H$\beta$ equivalent widths larger than
observed by almost a factor of ten.

\section{Discussion}

The analysis of emission line spectra of GEHR containing WR stars offer an
unique oportunity to test our present ideas about the evolution of star
forming regions and check on the accuracy of current stellar evolution
and atmosphere models. 
Most work on stellar population synthesis of
ionised regions assume that the ionising population responsible for the
observed gas emission belongs to a single star cluster created in a
unique star formation episode.
Therefore the WR emission features can be used to date these star
clusters which, in turn, should reproduce the observed emission line 
intensities if the other parameters controling the emission line
spectra, namely elemental abundances, particle density and ionisation
parameter, are known. Four of our observed HII regions meet all these 
requirements. 

\subsection{HII regions with observed WR features}

\subsubsection{H13 in NGC~628}

H13 in NGC~628, has a mean oxygen abundance of 12+log(O/H)= 8.24, about 0.2
times solar if  12+log(O/H)= 8.92 is assumed as the solar
value. The models of SV98 for this metallicity reproduce the observed
WR features for an ionising cluster of 4.1 Myr  which produces an H$\beta$
equivalent width of 159 {\AA}. In this cluster the proportion of WR to O
stars is WR/O = 5.1$\times$10$^{-2}$. Since both SV98 and STARBURST99 models make
use of the same stellar evolution prescriptions (
see Section 3) we have
assumed that a STARBURST99 model of a cluster of metallicity 0.2 times
solar which contains the same WR/O star ratio provides the spectral energy 
distribution of the ionising radiation. This model however provides an
[OIII]$\lambda$5007 {\AA} line emission which is higher than observed by a factor
of about 4. Also the [SII] and [NII] lines are down by a factor of
about 2 and the equivalent width of H$\beta$ is somewhat larger than
observed.

On the other hand, our best photoionisation
model corresponds to an age of 4.7 Myr producing an H$\beta$
equivalent width of 108 {\AA}.  The most appreciably discrepancy between
predicted and observed emission line ratios is found for the
[OII]$\lambda$3727 {\AA} line which is overestimated in the model by a factor of
30\% . This line is the one more affected by
reddening since the rest of the emission lines are measured relative to
a nearby hydrogen recombination line. The rest of the spectrum is
reproduced remarkably well (see Table 4), including the equivalent 
width of H$\beta$ 
(inside 30 \%) and the continuum luminosity at 9000  {\AA} (inside a factor
of two). Clusters younger than 4.7 Myr have a spectral energy
distribution which
results too hard to explain the observations. According to SV98 models
a cluster of this age would produce WR luminosities and equivalent
widths which are lower than observed by a factor of two, corresponding
to a WR/O ratio of 4.2$\times$10$^{-2}$. Given the
uncertainties involved in the different models employed, we can
conclude that a single star cluster with age between 4.0 and 4.7 Myr
fit all the observations adequately. The ionising spectrum of the
cluster of 4.7 Myr is shown in Fig. 2.

\subsubsection{CDT1 in NGC~1232}

Region CDT1 in NGC~1232 shows the highest oxygen abundance of the
regions with WR features ( 12+log(O/H)= 8.95, about solar if 
12+log(O/H)$_{\odot}$= 8.92). 
As explained above, no consistent solution for a single ionising
population is found to reproduce the observed WR features.  For this
region, two star clusters, with 2.3 and 5.7 Myr respectively,
provide Q(He)/Q(H) ratios nearly identical and corresponding to the 
stellar effective temperature of about 35000 K, in the Mihalas scale,
which reproduces the general features of the emission line spectrum
(see Paper I). Their ionising spectra are shown in Fig. 3.

The oldest of
the two clusters yields emission line ratios in better agreement with the
observations and an equivalent width of H$\beta$ closer to the observed one
(see Table 8). Clusters older than this cannot provide the necessary
photons to ionise the region.  SV98 models for
this age (5.7 Myr) give however WR features with luminosities relative to H$\beta$ which are a factor of about 10 larger than
observed.  On the other hand, the young 2.3 Myr cluster is more successful in reproducing
both the WR blue `bump' luminosity and equivalent width although
overpredicts the equivalent width of H$\beta$. These two facts taken
together point to the presence of a composite population with, at least
two clusters: a young one of about 2.5 Myr containing the WR stars and
an older one providing the bulk of the continuum light at $\lambda$ 4800 {\AA}.
We have therefore run a model using as ionising source the combination
of the spectral energy distributions 
of two ionising clusters of 2.4 and 7.1 Myr of age calculated with the
STARBURST99 code in which the youngest of
the two provides 10 times  the number of ionising photons emitted by
the oldest. This cluster contains the same WR/O  star ratio as
the SV98 model reproducing the observed WR features. The computed
photoionisation 
model is able to reproduce, with a reasonable degree of accuracy, most
of the observational constraints: the
emission line spectrum, the WR features both in luminosity and
equivalent width, the equivalent width of H$\beta$ and the continuum
luminosity at 9000 {\AA}. The characteristics of the two clusters, together
with the corresponding predictions of the photo-ionisation model are
given in Table 15. The particle density and the elemental abundaces
have been kept at their derived values given in Table 1.

\subsubsection{CDT3 in NGC~1232}

For region CDT3 in NGC~1232, the SV98 model which best reproduces the
observed WR features has an age of 3.1 Myr. The ratio of WR to O stars
is WR/O=5.4$\times$10$^{-2}$. A STARBURST99 model cluster with the same WR/O ratio
has 3.3 Myr of age and provides an H$\beta$ equivalent with of 159 {\AA} , close
to the observed one (189 {\AA}). However, when used as input to the
photoionisation model, this cluster provides [OIII]$\lambda$5007 {\AA} line
emission higher than observed by a factor of about 5. This is not
unexpected since the equivalent temperature corresponding to this
cluster is around 38200 K, while that estimated from photoionisation
models is only 34900 K (see Paper I). Clusters slightly older, also
reproducing the observed WR features within the errors, have larger
WR/O number star ratios (up to 8.3 $\times$10$^{-2}$) and therefore their
spectral energy distributions result even harder.

From the analysis of the emission line spectrum, two ionising clusters
are found with the same Q(He)/Q(H) number photon ratio corresponding to
the derived equivalent temperature of 34900 K (see Table 10 and Fig. 4). The
younger of these clusters however do not contain WR stars (WR/O=0.0)
and the oldest has WR/O=2.65$\times$10$^{-2}$, a factor of about 2 lower than
required to reproduce the WR features.

From the point of view of ionisation, the combination of two clusters
with ages 2.8 and 4.8 Myr contributing approximately 90 \% and 10 \% to
the total number of ionising photons would reproduce the emission line
spectrum rather well, except for the [OII]3727 {\AA} line which results
about 40\% higher in the model. Also the equivalent width of H$\beta$ and
the continuum light at $\lambda$9000 {\AA} are well reproduced. However, for this
population, the luminosity and equivalent width of the WR blue `bump'
are  lower than observed by factors of about 15 and 25 respectively.

\subsubsection{CDT4 in NGC~1232}

A similar situation is found for region CDT4. The SV98 model which best
reproduces the observed WR features has an age of 4.1 Myr. The ratio of
WR to O stars is WR/O=9.2$\times$10$^{-2}$, the highest of all the observed
regions. The STARBURST99 model cluster of the same age also has the
same WR/O ratio
and provides an H$\beta$ equivalent width of 131 \AA , that should be compared
with the observed one of 138 {\AA}. When introduced as input in the
photoionisation code, this cluster reproduces the emission line
spectrum satisfactorily, except for the  [OIII]$\lambda$5007 {\AA} line
which is higher than observed by a factor of about 3. Again this is
related to the hardness of the ionising continuum which provides a
Q(He)/Q(H) ratio of 0.18 which corresponds to an equivalent temperature
of 37800 K while the corresponding one derived from the emission line
spectrum is only 35000 K (see Paper I). 

Again, the analysis of the emission line spectrum indicates two
possible star clusters as ionising sources whose ages coincide with
those found for region CDT3 (see Table 11 and Fig. 4). Therefore, regarding the ionisation, the
same discussion above applies to this region: it is possible to find a
combination of ionising clusters which reproduces well the emission
line spectrum, but predicts WR feature luminosities and equivalent
widths well below the observed values. 
 
\subsection{HII regions with no WR features observed}

In the rest of the regions no WR features are observed and therefore
the estimation of the age of the ionising populations must be made with
the only help of the analysis of the emission line spectra. Out of the two
solutions found for region H3 in NGC628, the cluster with 4.5 Myr
provides the best fit to the emission line spectrum (Table 5). Most line ratios
are predicted to better than 10 \%. The predicted H$\beta$ equivalent width
however, is 60 \% lower than observed.  On the other hand, this cluster
has a WR/O number star ratio of 5.8$\times$10$^{-2}$, which would provide a
WR blue `bump'
 luminosity relative to H$\beta$ of 0.036 and an equivalent
width of 3.78 {\AA}, that should be observable.  The younger cluster, of 2.9 Myr, does not have WR stars
and, although the fit to the emission line spectrum is slightly worse,
it does predict adequately both the equivalent width of H$\beta$ and the
continuum luminosity at $\lambda$ 9000 {\AA}. Therefore, a single young ionising
cluster of less than 3 Myr, in an evolutionary state prior to the
appearance of WR stars produces a satisfactory fit to all the
observational constraints. The spectral energy distribution of the 2.9
Myr cluster is shown in Fig. 2.

The ionisation of regions H4 and H5 also requires young clusters without WR
stars.
Clusters of 2.5 Myr and 2.9 Myr respectively  reproduce satisfactorily
the emission line
spectra, the H$\beta$ equivalent widths and the luminosity at 9000 {\AA} (see
Tables 6 and 7). The cluster in H5 seems to be slightly older than that
in H4 and the ionisation parameter in this region is lower by a
factor of about 2. The SEDs of the two ionising clusters is shown in
Fig. 5.

%%%%%%%%%%%%%%%%%%%%%%%%%%%%%%%%%%%%%%%%%%%%%%%%%%%%%%%%%%%%%%%%%%%%%%%%%%%%%%%%%%%

The region in NGC~1232 not showing WR features, CDT2, also allows for
two possible ionising clusters (see Table 9 and Fig. 6). 
According to the STARBURST99 models, the
oldest one with 4.5 Myr has a WR/O star ratio of 0.1 which, in 
SV98 models, gives WR
blue `bump' luminosity and equivalent width of 0.1 and 6.8 {\AA}
respectively. On the other hand, the youngest cluster with 3 Myr, has a
WR/O star ratio of only 0.01 and provides  WR
blue `bump' luminosity relative to H$\beta$ and equivalent width of 0.01
and 2.34 {\AA} respectively. Based
on the fact that no WR features are observed, we consider this latter
cluster as a better candidate for the ionisation of the
region. Nevertheless, it does not seem to be the only population
present since the low observed value of the H$\beta$ equivalent width
points to an underlying population that would contribute 70 \% to the
luminosity at H$\beta$. The WR blue `bump' equivalent width would be
consequently lowered to only 0.7 {\AA} which is consistent with the lack of
detection of the feature.

The other region of the sample showing a high metallicity (about twice
solar) is CDT1 in NGC~1637. Two ionising clusters with Z=0.04 are 
able to reproduce the
observed emission line spectrum to the same degree of accuracy and they
produce H$\beta$ equivalent widths higher and lower than observed
respectively (see Table 12 and Fig. 7)). Also the observed $\lambda$9000 {\AA} 
luminosity is intermediate to
that predicted by the two clusters. Therefore, the best solution would
seem to consist of a combination of the two. We should however keep in
mind that, according to models, both clusters contain WR stars with
WR/O ratios between 0.14 and 14. The oldest cluster would provide
large values of both blue `bump' luminosity and equivalent width
(0.7 and 13 {\AA}) while the younger one would provide about the same
equivalent width but a luminosity lower by a factor of about 5
(0.14). Therefore, based on the fact that no WR features are observed
we feel that the most plausible solution would be the combination of a
young ionising cluster providing the bulk of the ionisation and an
underlying population which would lower the equivalent width of H$\beta$ to
the observed values. This underlying population should contribute 50\%
of the continuum luminosity at $\lambda$ 4861~{\AA}.

Finally, not much can be said about the two regions in NGC~925. Both of
them show a metallicity of about 12+log(O/H)$\simeq$8.5, although, due to
the non detection of suitable temperature sensitive lines, the
uncertainty in the derived oxygen abundance is high ($\pm$ 0.20). In the
case of CDT1, the emission line
spectrum  can be adequately reproduced by relatively old ionising
clusters of slightly different age  and metallicity (5.8 Myr and
12+log(O/H) = 8.52 or 5.9 Myr and 12+log(O/H) = 8.62) with the best
solution probably being in between (see Table 13). There is
however some room for contribution by a non ionising cluster since 
the luminosity at $\lambda$9000 {\AA} is predicted to be lower than
observed by a factor of about 3. 

In the case
of CDT4, the 5.9 Myr cluster predicts an equivalent width of H$\beta$ lower 
than
observed by a factor of about two, while a young cluster without WR stars
(2.8 Myr) gives a value much larger than observed (Table 14). It is however
possible to find a
satisfactory solution by considering the combination of two clusters of
2.8 and 5.9 Myr with this former  one providing 60\%  of the total number  of
ionising photons. This solution is also listed in Table 14.

\subsection{Global analysis}

In summary, out of the 11 analysed regions, only the four in
NGC~628  seem to be
consistent with the presence of single ionising star clusters, all of
them show relatively low abundances. For the rest of the
regions, at least two populations are needed: a young one, with age
between 2.5 and 3 Myr, which
provides most of (or all) the ionising photons, and an older one which
contributes a substantial fraction of the continuum luminosity. This
second population also contains an important fraction of the mass of
the composite cluster. The difference in age between the two
populations is not very large, 3-10 Myr. 

Composite populations for HII regions have been found in previous
work by Mayya \& Prabhu (1996) and by D{\'\i}az et al. (2000b) for disc and
circumnuclear objects respectively from broad band and H$\alpha$ photometry.
A spectrophotometric analysis of circumnuclear HII regions in NGC~3310
(Pastoriza  et al. 1993) and NGC~7714 (Garc{\'\i}a-Vargas et al. 1997) also
reveals the presence of composite populations and even in the young
HII region 30 Dor in the LMC a spread in 
age for the stellar populations has been found (Selman et al. 1999).

This scenario, is able to reproduce both the emission line spectrum and
the WR features of the high metallicity region CDT1 in NGC~1232. It is
nevertheless not sufficient to explain the
observations in the two intermediate metallicity HII regions in
NGC~1232 with WR features: CDT3 and CDT4. For these two regions, no consistent
solution can be found. While the models by SV98 consistently provide
excellent fittings to the observed WR properties, the evolutionary
models by STARBURST99 result too hard to explain the observed emission line
spectra. 

Several solutions can be suggested such as the inadequacy of 
the stellar evolutionary tracks, uncertainties in the computation of
stellar atmosphere models or the everlasting question of
temperature fluctuations within these regions that would underestimate
the true abundances.

The fact that the WR features are adequately reproduced by SV98 models
seem to imply that the evolutionary tracks are able to predict the
right relative numbers of WR/O stars, and their different subtypes at the
derived abundances. These relative numbers, combined with the observed 
emission line luminosities of the individual WR stars, and the predicted
continuum energy distribution of the ionising population predict
emission line intensities and equivalent widths of the WR stars that are in
excellent agreement with observations. This can be seen in Fig 9. Upper
and lower  panels respectively  show the predicted emission line
intesities and equivalent widths of the WR 
blue `bump' for different metallicities: 0.2 solar, 0.4 solar and
solar together with the data corresponding to the four observed
objects. Solid symbols correspond to the observed data. In the case of CDT1 in NGC~1232, for which a composite
population is found, two open symbols are shown, one corresponding to
the predictions for a single ionising cluster, and another one
including the contribution to the H$\beta$  and continuum luminosity of
the cluster not containing WR stars. For regions CDT3 and CDT4
observations and predictions are nearly identical for the respective
clusters and therefore only one symbol is shown.  As we can see, the
agreement for the three regions in NGC~1232 is surprisingly good. 
For region H13 in
NGC~628 predicted and observed values (solid and open symbols) agree
inside a factor of about two. We therefore conclude that, 
regarding the WR star populations, the currently assumed  stellar
evolution, including a high mass loss, is adequate to match the
observations, at least for intermediate
and high metallicities. The worse agreement found
for region H13 in NGC~628, whose WR feature luminosity and equivalent
width seem to indicate a slightly higher metallicity, might in principle be
attributed to the presence of temperature fluctuations, which do not
need to be invoked in the other cases.

The third possible source of disagreement concerns the existing
uncertainties in
modelling the atmospheres of high mass stars of O and WR types. Our
results seem to point to an overestimate of the hardness of the
spectral energy distribution of the stars in the WR phase which is more
apparent for intermediate to high metallicities. The same sort of
effect has been found by DCTG00 for region 74C in NGC~4258
and Esteban et al. (1993) for the galactic WR
nebula M1-67. These latter authors, from stellar and nebular
spectroscopic analyses,
concluded that lower temperatures were required from the
photoionisation models for late type WN (WNL) stars. In their study
they used the unblanketed WR models of Schmutz et al. (1992). A
subsequent reanalysis of this region has been made by Crowther et al.
(1999) using
blanketed model atmospheres. In this case the resulting ionising
spectrum is much softer and reproduces better the observations although
some important discrepancies, like a negligible
HeI $\lambda$ 5876 {\AA} line intensity, which is nonetheless observed, 
and the overprediction of [OII] $\lambda$ 3727 {\AA} and [NII]  
$\lambda$ 6584 {\AA} by factors of 3 and 2 respectively, still remain.

It should however be recalled that all results discussed above are
based on ionisation
bounded models. An interpretation of the emission line spectra of
HII region with WR features on the basis of matter bounded models has
been presented in Castellanos, D{\'\i}az \& Tenorio-Tagle (2002). Models of 
this kind have
been found to provide excellent fittings to the observations of H13 in
NGC~628, CDT3 in NGC~1232 and 74C in NGC~4258 using both SV98 and
STARBURST99 models. These models would point to an important leakage
of ionising photons depending on both the metallicity and evolutionary
state of the region.

Certainly, the inclusion of line
blanketing in the  model atmospheres of WR stars, which is already
included in the CoStar (Schaerer \& de Koter 1997) and Lejeune et
al. (1997) models is imperative to interpret correctly the emission
line spectra observed in HII regions.

%%%%%%%%%%%%%%%%%%%%%%%%%%%%%%%%%%%%%%%%%%%%%%%%%%%%%%%%%%%%%%%%%%%%%%%%%%%%%%

%        CONCLUSIONES
%%%%%%%%%%%%%%%%%%%%%%%%%%%%%%%%%%%%%%%%%%%%%%%%%%%%%%%%%%%%%%%%%%%%%%%%%%%%%%

\section{Summary and conclusions} 

We have analysed the stellar populations of eleven  H{\sevensize II} regions 
in the galaxies NGC 628, 
NGC 1232, NGC 925 and NGC 1637 using spectrophotometric observations 
between 3500 and 9700 {\AA}.  We have  derived the physical 
properties of the regions and their corresponding ionising clusters: 
filling factor, mass of ionised gas and mass of ionising stars. 
Most of the regions have small ionising clusters with masses in the
range 1500 to 
30000 solar masses. The exceptions are the three supergiant HII 
regions in NGC 1232, CDT1, CDT3 and CDT4 with masses greater than 
100,000 solar masses. These values constitute in fact lower limits since the 
regions are assumed to be ionisation bounded and the presence of 
dust has not been taken into account.

WR features have been observed in the four supergiant HII regions,  H13
in NGC~628 and CDT1, CDT3, CDT4 in NGC~1232.  
A very detailed modelling has been carried out for these regions by using
two different sets 
of models: those of SV98 for WR star populations and those of 
Leitherer et al. (1999) for 
ionising populations, to try to reproduce simultaneously both the WR features 
and the emission line spectrum of each region. In all cases, the agreement between both 
predicted and observed values for the WR luminosities and equivalent widths
is excellent. This fact seems to indicate that the stellar evolution
assumed by the models is predicting the right WR/O ratios at the
different metallicities involved.

The ages of the populations containing WR stars are found
to be between 2.3 and 4.1 Myr. 
This population is sufficient to explain the emission line spectrum of
the low metallicity region H13 in NGC~628, taking into account the
uncertainties involved in both observations and model computations. 

This is not the case for the rest of the regions. For CDT1 in NGC~1232,
a second ionising population is required to simultaneously reproduce
both the WR
features and the emission line spectrum. This second population is
older than the previous one by about 5 Myr and contributes about 10\% of
the ionising photons and 5 times the continuum luminosity at 9000 {\AA}.
Composite populations are also found for half of the regions without WR
features, although in this case, the result is based only on the
emission line spectrum analysis.

For the other two regions containing WR stars,
CDT3 and CDT4 in NGC~1232, no consistent solution is found, since the
population containing WR stars produces a spectral energy distribution
which is too hard to explain the emission of the gas. Several solutions
are proposed to solve this problem. Both a reduction in the number of
high energy photons of the ionising clusters, which could arise
naturally with the inclusion of
blanketing by metallic lines in the atmospheres of the WR stars, and/or
the assumption that the HII regions be matter bounded can effectively
solve the problem.
The latter hypothesis has been tested successfully (Castellanos et
al. 2002) while the former one still remains to be explored.

\section{Acknowledgements}

We would like to thank L. Van Zee for her rapid and helpful response
when required. We also thank an anonymous referee for several useful comments. E.T. is grateful to an IBERDROLA Visiting Professorship 
to UAM during which part of this work was completed and to the Mexican research council (CONACYT) for support through the research grant \# 211290-5-32186E.

This work has been partially supported by MCyT project AYA-2000-093.

%%%%%%%%%%%%%%%%%%%%%%%%%%% Referencias %%%%%%%%%%%%%%%%%%%%%%%%%%%%%%%%%%%%%%%%

%%%%%%%%%%%%%%%%%%%%%%%%%% Final de las referencias %%%%%%%%%%%%%%%%%%%%%%%%%%%

%%%%%%%%%%%%%%%%%%%%%%%%%%%%TABLAS%%%%%%%%%%%%%%%%%%%%%%%%%%%%%%%%%%%%%%%%%%%%%

\newpage

%%%%%%%%%%%%%%%%%%%%%%%%%%%%%%%%%%%%%%%%%%%%%%%%%%%%%%%%%%%%%%%
%                                                             %
%             Tabla 1 Summary of observations                 % 
%%%%%%%%%%%%%%%%%%%%%%%%%%%%%%%%%%%%%%%%%%%%%%%%%%%%%%%%%%%%%%%

\begin{table*}
\setcounter{table}{0}
 %\begin{minipage}{180mm}
 \caption{Summary of derived gas physical conditions of the analysed HII regions}
 \begin{tabular}{llccccccccc}
Galaxy & Region & n$_e$ & t(S$^{++}$) & t(N$^{+}$) & $<$t$>_{adopt}$ & 12+log(O/H) & log
(N/O) & log(S/O) & log U & T$_{eff}$(K) \\
         &              &             &             &      &          &
             &              &              &       \\
          
NGC~628  & H13  & 80    & 1.02$\pm$0.03 & 0.90$\pm$0.06 & --   & 8.24$\pm$0.08 &
-1.08$\pm$0.04 & -1.83$\pm$0.04 & -2.78$\pm$0.10 & 35000 \\
         & H3   & $\leq$40 & 1.03$\pm$0.05 & --          & --   & 8.23$\pm$0.15 &
-1.14$\pm$0.10 & -1.85$\pm$0.11 & -2.92$\pm$0.15 & 36000 \\
         & H4   & $\leq$40 &  0.95$\pm$0.05 & --          & --   & 8.31$\pm$0.15 &
-0.97$\pm$0.10 & -1.80$\pm$0.11 & -2.95$\pm$0.15 & 35000 \\
         & H5   & $\leq$40 & 0.82$\pm$0.06 & --          & --   & 8.34$\pm$0.15 &
-1.07$\pm$0.10 & -1.67$\pm$0.10 & -2.97$\pm$0.15 & 34800 \\
         &              &             &             &      &          &
             &              &              &       \\
NGC~925  & CDT1 & $\leq$40 & --          & --          & 0.86 & 8.52  &
-1.07        & -1.83        & -3.00$\pm$0.30 & 36500 \\
         & CDT4 & $\leq$40 & --          & --          & 0.93 & 8.41  &
-1.01        & -1.83        & -3.00$\pm$0.25 & 36000 \\ 
         &              &             &             &      &          &
             &              &              &       \\
NGC~1232 & CDT1 & 130   & 0.54$\pm$0.05 & 0.67$\pm$0.05 & --   & 8.95$\pm$0.20 &
-0.81$\pm$0.08 & -1.81$\pm$0.10 & -2.95$\pm$0.20 & 34900 \\
         & CDT2 & $\leq$40 & --          & --          & 0.81 & 8.61$\pm$0.15 &
-1.15$\pm$0.10 & -1.81$\pm$0.15 & -2.95$\pm$0.15 & 36900 \\
         & CDT3 & 223   & 0.74$\pm$0.05 & 0.86$\pm$0.06 & --   & 8.56$\pm$0.14 & 
-0.96$\pm$0.04 & -1.70$\pm$0.07 & -2.72$\pm$0.10 & 34900 \\
         & CDT4 & 118   & 0.87$\pm$0.04 & 0.90$\pm$0.06 & --   & 8.37$\pm$0.12 &  
-0.91$\pm$0.06 & -1.62$\pm$0.07 & -2.72$\pm$0.10 & 35000 \\
         &              &             &             &      &          &
             &              &              &       \\
NGC~1637 & CDT1 & 100   & --          & --          & 0.40 & 9.10  &
-0.88        & -1.97        & -3.30$\pm$0.20 & 35000 \\
             & & & & & & & & & & \\
             & & & & & & & & & & \\ 
            & & & & & & & & & & \\
 \end{tabular}
 %\end{minipage}
\end{table*}

%%%%%%%%%%%%%%%%%%%%%%%%%%%%%%%%%%%%%%%%%%%%%%%%%%%%%%%%%%%%%%%%%%%%%%%%%%%
%%%%%%%%%%%%%%%%%%%%%%%%%%%%%%%%%%%%%%%%%%%%%%%%%%%%%%%%%%%%%%%%%%%%%
%                                                                   %                
% Tabla 2: Propiedades fisicas de las regiones HII observadas       %
%                                                                   %
%                                                                   %
%%%%%%%%%%%%%%%%%%%%%%%%%%%%%%%%%%%%%%%%%%%%%%%%%%%%%%%%%%%%%%%%%%%%%

\begin{table*}
\setcounter{table}{1}
%\begin{minipage}{120mm}
 \caption{Physical properties of the observed HII regions}
 \begin{tabular}{llcccccc}
Galaxy & Region & L(H$\alpha$) & Q(H) & $\epsilon$ & M(HII) & M$^{\star}$ & $\phi$ \\
       &        & (10$^{38}$ erg s$^{-1}$) & (10$^{49}$ s$^{-1}$) & &
       (M$_{\odot}$) & (M$_{\odot}$) & 
(arcsec)\\
        &       &         &          &        &       &        &     \\
NGC~628 & H13   &  6.63   & 48.5     & 0.05   & 17900 & 22912  & 1.8 \\
        & H3    &  1.19   & 8.73     & 0.11   & 25800 &  2680  & 1.3 \\
        & H4    &  0.695  & 5.09     & 0.13   & 15000 &  1720  & 1.0 \\
        & H5    &  0.81   & 5.88     & 0.11   & 17300 &  1920  & 1.1 \\
        &       &         &          &        &       &        &     \\
NGC~925 & CDT1  &  0.91   & 6.67     & 0.41   & 19700 & 15475  & 1.5 \\
        & CDT4  &  1.27   & 9.27     & 0.25   & 27400 & 11196  & 2.0 \\
        &       &         &          &        &       &        &     \\
NGC~1232& CDT1  &  20.8   & 152      & 0.008  & 20800 & 180290 & 1.0 \\
        & CDT2  &  2.27   & 16.6     & 0.07   & 49000 & 12168  & 0.6 \\
        & CDT3  &  38.5   & 282      & 0.015  & 25200 & 102900 & 0.9 \\
        & CDT4  &  48.2   & 353      & 0.02   & 88300 & 168841 & 1.3 \\
        &       &         &          &        &       &        &     \\
NGC~1637& CDT1  & 1.77    & 13.0     & --     &  --   & 10586  & 1.5 \\
        &       &         &          &        &       &        &     \\
        &       &         &          &        &       &        &     \\
        &       &         &          &        &       &        &     \\
 \end{tabular}
 %\end{minipage}
\end{table*}

%%%%%%%%%%%%%%%%%%%%%%%%%%%%%%%%%%%%%%%%%%%%%%%%%%%%%%%%%%%%%%%%%%%%%%%%%%%%
% Tabla 3: WR feature intensities and equivalent widths in the observed
% regions
%%%%%%%%%%%%%%%%%%%%%%%%%%%%%%%%%%%%%%%%%%%%%%%%%%%%%%%%%%%%%%%%%%%%%%%%%%%%

\begin{table*}
\setcounter{table}{2}
%\begin{minipage}{150mm}
 \caption{ WR feature intensities and equivalent widths in the observed
   HII regions.}
 \begin{tabular}{lcccc}
Region & L(WR)/H$\beta$ & EW(WR)({\AA} ) & L(HeII)/H$\beta$ & EW(HeII)({\AA} ) \\
Region H13 (NGC 628)   & 0.08 &  8.9 & 0.04 & 4.8 \\ 
Region CDT1 (NGC 1232) & 0.04 &  2.0 & 0.015 & 0.8 \\
Region CDT3 (NGC 1232) & 0.03 &  5.9 & 0.015 & 2.7 \\
Region CDT4 (NGC 1232) & 0.08 &   10 & 0.03 & 3.9 
\end{tabular}
 %\end{minipage}
 \end{table*}
%%%%%%%%%%%%%%%%%%%%%%%%%%%%%%%%%%%%%%%%%%%%%%%%%%%%%%%%%%%%%%%%%%%%%

\clearpage

%%%%%%%%%%%%%%%%%%%%%%%%%%%%%%%%%%%%%%%%%%%%%%%%%%%%%%%%%%%%%%%%%%%%%
%                                                                   %
%                                                                   %
%                                                                   %
% Tabla 4 a 13: Modelos de cumulos ionizantes que mejor ajustan las %
% observaciones                                                     %
%                                                                   %
%                                                                   %
%%%%%%%%%%%%%%%%%%%%%%%%%%%%%%%%%%%%%%%%%%%%%%%%%%%%%%%%%%%%%%%%%%%%%

%%%%%%%%%%%%%%%%%%%%%%%%%%%%%%%%%%%%%%%%%%%%%%%%%%%%%%%%%%%%%%%%%%%%%%%%%%%%%%%
%
%       REGION H13 EN NGC 628 (Z = 0.2 SOLAR + WR FEATURES)
%
%%%%%%%%%%%%%%%%%%%%%%%%%%%%%%%%%%%%%%%%%%%%%%%%%%%%%%%%%%%%%%%%%%%%%%%%%%%%%%%

\begin{table}
\setcounter{table}{3}
 %\begin{minipage}{60mm}
  \caption{Evolutionary models for region H13 in NGC 628}
  \begin{tabular}{@{}lcccc@{}}
   Parameter & Observations & 4.7 Myr \\
             &              &         \\
 Q(He)/Q(H)  &              & 0.088 \\
 Q(He$^+$)/Q(He)&              & 1.1$\times$10$^{-4}$ \\
 WR/O        &              & 0.042\\      
 $<$log U$>$ & -2.78 $\pm$ 0.10 & -2.88 \\
 n$_{e}$ & 80 & 100 \\  
 12 + log(O/H) & 8.24 $\pm$ 0.08 & 8.24 \\
 log(S/O) & -1.83 $\pm$ 0.04 & -1.78 \\
 log(N/O) & -1.08 $\pm$ 0.04 & -1.11 \\
 3727 [OII]  & 2960 $\pm$ 110 & 3852 \\
 5007 [OIII] & 1547 $\pm$ 10 & 1530 \\
 4363 [OIII] & 10 $\pm$ 2 & 10 \\
 6584 [NII]  & 496 $\pm$ 8 & 514 \\
 5755[NII]   & 5 $\pm$ 1 & 9 \\
 6717 [SII]  & 204 $\pm$ 5 & 210 \\
 4072[SII]   & 29 $\pm$ 3 & 30 \\ 
 9069 [SIII] & 168 $\pm$ 10 & 179 \\
 6312 [SIII] & 12 $\pm$ 1 & 11 \\
 t(O$^{2+}$) & 0.98 $\pm$ 0.05 & 0.99 \\
 t(S$^{2+}$) & 1.02 $\pm$ 0.03 & 1.02 \\
 t(S$^{+}$)  & 0.99 $\pm$ 0.06 & 0.98 \\
 t(N$^{+}$)  & 0.90 $\pm$ 0.06 & 1.04 \\
 EW(H$\beta$)({\AA}) & 140 & 108 \\
L$_{9000}$ (10$^{35}$erg s$^{-1}${\AA}$^{-1}$) & 2.9 & 5.7 
\end{tabular}
%\end{minipage}
\end{table}

%%%%%%%%%%%%%%%%%%%%%%%%%%%%%%%%%%%%%%%%%%%%%%%%%%%%%%%%%%%%%%%%%%%%%%%%%%%%%%%
%
% REGION H3 EN NGC 628 (Z = 0.2 SOLAR)
%
%%%%%%%%%%%%%%%%%%%%%%%%%%%%%%%%%%%%%%%%%%%%%%%%%%%%%%%%%%%%%%%%%%%%%%%%%%%%%%%

\begin{table}
\setcounter{table}{4}
 %\begin{minipage}{60mm}
  \caption{Evolutionary models for region H3 in NGC 628}
  \begin{tabular}{@{}lccc@{}}
   Parameter & Observations & 2.9 Myr \\
             &              &         \\
Q(He)/Q(H)  &              & 0.13 \\
 Q(He$^+$)/Q(He)&           & 0.00 \\
 WR/O        &              & 0.0          \\   
 $<$log U$>$ & -2.92 $\pm$ 0.15 & -2.89 \\
 n$_e$        &  $\leq$40          & 10  \\  
 12 + log(O/H) & 8.23 $\pm$ 0.15 & 8.23  \\
 log(S/O) & -1.85 $\pm$ 0.11 & -1.73  \\
 log(N/O) & -1.14 $\pm$ 0.10 & -1.11   \\
 3727 [OII]  & -- & 4569  \\
 5007 [OIII] & 1596 $\pm$ 12 & 1546  \\
 4363 [OIII] & 16 $\pm$ 2 & 11  \\
 6584 [NII]  & 470 $\pm$ 20 & 484  \\
 6717 [SII]  & 270 $\pm$ 13 & 234  \\
 9069 [SIII] & 136 $\pm$ 10 & 219  \\
 6312 [SIII] & 11 $\pm$ 1 & 14  \\
 t(O$^{2+}$) & 1.17 $\pm$ 0.10 & 1.03  \\
 t(S$^{2+}$) & 1.03 $\pm$ 0.05 & 1.06  \\
 EW(H$\beta$)({\AA}) & 231 & 222  \\
L$_{9000}$ (10$^{34}$erg s$^{-1}${\AA}$^{-1}$) & 5.6 & 6.2 
\end{tabular}
%\end{minipage}
\end{table}

\clearpage

%%%%%%%%%%%%%%%%%%%%%%%%%%%%%%%%%%%%%%%%%%%%%%%%%%%%%%%%%%%%%%%%%%%%%%%%%%%%%%%
%
%       REGIONES H4 EN NGC 628 (Z = 0.2 SOLAR)
%
%%%%%%%%%%%%%%%%%%%%%%%%%%%%%%%%%%%%%%%%%%%%%%%%%%%%%%%%%%%%%%%%%%%%%%%%%%%%%%%

\begin{table}
\setcounter{table}{5}
 %\begin{minipage}{60mm}
  \caption{Evolutionary models for region H4 in NGC 628}
  \begin{tabular}{@{}lccc@{}}
   Parameter & Observations (H4) & 2.5 Myr \\
             &                   &         \\
 Q(He)/Q(H)  &                   & 0.18  \\
 Q(He$^+$)/Q(He)&                 & 0.00  \\
 WR/O        &                   & 0.0   \\ 
 $<$log U$>$ & -2.95 $\pm$ 0.15 & -2.95 \\
 n$_{e}$ & $\leq$40 & 10  \\  
 12 + log(O/H) & 8.31 $\pm$ 0.15 & 8.18 \\
 log(S/O) & -1.80 $\pm$ 0.11 & -1.71 \\
 log(N/O) & -0.97 $\pm$ 0.10 & -1.06 \\
 3727 [OII]  & -- & 4432 \\
 5007 [OIII] &  2177 $\pm$15 & 2000 \\
 6584 [NII]  & 527 $\pm$ 20 & 500 \\
 6717 [SII]  & 308 $\pm$ 20 & 245 \\
 9069 [SIII] & 147 $\pm$ 12 & 222 \\
 6312 [SIII] & 10 $\pm$ 2   & 16 \\
 t(O$^{2+}$) & 0.94 & 1.09 \\
 t(S$^{2+}$) & 0.95 $\pm$ 0.05 & 1.11 \\
 EW(H$\beta$)({\AA}) & 220 & 290 \\
L$_{9000}$ (10$^{34}$erg s$^{-1}${\AA}$^{-1}$) & 2.9 & 4.2
\end{tabular}
%\end{minipage}
\end{table}

%%%%%%%%%%%%%%%%%%%%%%%%%%%%%%%%%%%%%%%%%%%%%%%%%%%%%%%%%%%%%%%%%%%%%%%%%%%%%%%
%
%       REGIONES H5 EN NGC 628 (Z = 0.2 SOLAR)
%
%%%%%%%%%%%%%%%%%%%%%%%%%%%%%%%%%%%%%%%%%%%%%%%%%%%%%%%%%%%%%%%%%%%%%%%%%%%%%%%

\begin{table}
\setcounter{table}{6}
 %\begin{minipage}{60mm}
  \caption{Evolutionary models for region H5 in NGC 628}
  \begin{tabular}{@{}lcccc@{}}
   Parameter &  Observations (H5) & 2.95 Myr \\
             &                    &          \\
 Q(He)/Q(H)    &                &  0.13 \\
 Q(He$^+$)/Q(He)&                &  0.00 \\
 WR/O          &                &  0.0  \\ 
 $<$log U$>$   & -2.97 $\pm$ 0.15 & -3.20 \\
 n$_{e}$         & $\leq$40          & 10  \\  
 12 + log(O/H) & 8.34 $\pm$ 0.15  & 8.36 \\
 log(S/O)      & -1.67 $\pm$ 0.10 & -1.71 \\
 log(N/O)      & -1.07 $\pm$ 0.10 & -1.16 \\
 3727 [OII]    &  --            & 5461 \\
 5007 [OIII]   &  529 $\pm$ 8     & 692 \\
 6584 [NII]    &  631 $\pm$ 43    & 750 \\
 6717 [SII]    &  391 $\pm$ 24    & 400 \\
 9069 [SIII]   &  149 $\pm$ 10    & 213 \\
 6312 [SIII]   &    7 $\pm$ 2     & 11 \\
 t(O$^{2+}$)     &     0.79        & 0.97 \\
 t(S$^{2+}$)     &   0.82 $\pm$ 0.06 & 1.0  \\
 EW(H$\beta$)({\AA})   &     205         & 222 \\
L$_{9000}$ (10$^{34}$erg s$^{-1}${\AA}$^{-1}$) & 3.4 & 4.2
\end{tabular}
%\end{minipage}
\end{table}

\clearpage

%%%%%%%%%%%%%%%%%%%%%%%%%%%%%%%%%%%%%%%%%%%%%%%%%%%%%%%%%%%%%%%%%%%%%%%%%%%%%%%
%
% REGION CDT1 EN NGC 1232 (ALTA MET + WR FEATURES)
%
%%%%%%%%%%%%%%%%%%%%%%%%%%%%%%%%%%%%%%%%%%%%%%%%%%%%%%%%%%%%%%%%%%%%%%%%%%%%%%%

\begin{table}
\setcounter{table}{7}
 %\begin{minipage}{60mm}
  \caption{Evolutionary models for region CDT1 in NGC 1232}
  \begin{tabular}{@{}lcccc@{}}
   Parameter & Observations & 2.3 Myr       & 5.7 Myr \\
             &              &               &         \\
 Q(He)/Q(H)  &              &  0.10         & 0.098  \\
 Q(He$^+$)/Q(He)&            & 3.5$\times$10$^{-5}$  & 1.8$\times$10$^{-4}$ \\
 WR/O        &              & 0.03          &  $>$0.48 \\ 
 $<$log U$>$ & -2.95 $\pm$ 0.20 & -2.95 & -3.10 \\
 n$_{e}$ & 130 & 130 & 130 \\  
 12 + log(O/H) & 8.95 $\pm$ 0.20 & 9.05 & 9.05 \\
 log(S/O) & -1.81 $\pm$ 0.10 & -1.81 & -1.81 \\
 log(N/O) & -0.81 $\pm$ 0.08 & -0.81 & -0.81 \\
 3727 [OII]  & 1490 $\pm$ 70 & 2190 & 1422 \\
 5007 [OIII] & 229 $\pm$ 4 & 263 & 239 \\
 5755 [NII]  & 4 $\pm$ 1 & 7 & 5 \\
 6584 [NII]  & 1040 $\pm$ 21 & 1600 & 1288 \\
 6717 [SII]  & 357 $\pm$ 8 & 340 & 387 \\
 6312 [SIII] & 2 $\pm$ 1 & 4 & 2 \\
 9069 [SIII] & 197 $\pm$ 13 & 291 & 170 \\
 t(N$^{+}$)  & 0.67 $\pm$ 0.06 & 0.63 & 0.58 \\
 t(O$^{2+}$) & 0.45 & 0.56 & 0.49 \\
 t(S$^{2+}$) & 0.54 $\pm$ 0.05 & 0.60 & 0.53 \\
 EW(H$\beta$)({\AA}) & 48 & 328 & 31 \\
 L$_{9000}$ (10$^{36}$erg s$^{-1}${\AA}$^{-1}$) & 4.2 & 0.9 & 4.7 
\end{tabular}
%\end{minipage}
\end{table}

%%%%%%%%%%%%%%%%%%%%%%%%%%%%%%%%%%%%%%%%%%%%%%%%%%%%%%%%%%%%%%%%%%%%%%%%%%%%%%%
%
%         REGION CDT2 EN NGC 1232 (Z = 0.5 SOLAR)
%
%%%%%%%%%%%%%%%%%%%%%%%%%%%%%%%%%%%%%%%%%%%%%%%%%%%%%%%%%%%%%%%%%%%%%%%%%%%%%%%

\begin{table}
\setcounter{table}{8}
 %\begin{minipage}{60mm}
  \caption{Evolutionary models for region CDT2 in NGC 1232}
  \begin{tabular}{@{}lcccc@{}}
   Parameter & Observations & 3.0 Myr & 4.5 Myr \\
             &              &         &         \\
 Q(He)/Q(H)  &    & 0.155  & 0.130  \\
 Q(He$^+$)/Q(He)&     & 5.4$\times$10$^{-4}$  & 0.062  \\
 WR/O        &       &   0.01      &  0.1        \\ 
 $<$log U$>$ & -2.95 $\pm$ 0.15 & -3.05 & -3.10 \\
 n$_{e}$ & $\leq$40 & 10 & 10 \\  
 12 + log(O/H) & 8.61 $\pm$ 0.15 & 8.60 & 8.55 \\
 log(S/O) & -1.81 & -1.71 & -1.71 \\
 log(N/O) & -1.15 & -1.15 & -1.15 \\
 3727 [OII]  & 4180 $\pm$ 220 & 4313 & 4026 \\
 5007 [OIII] & 1350 $\pm$ 20 & 1375 & 1334 \\
 6584 [NII]  & 717 $\pm$ 30 & 710 & 694 \\
 6717 [SII]  & 537 $\pm$ 15 & 517 & 523 \\
 9069 [SIII] & 191 $\pm$ 15 & 294 & 257 \\
 t(O$^{2+}$) & -- & 0.79 & 0.81 \\
 t(S$^{2+}$) & -- & 0.83 & 0.84 \\
 EW(H$\beta$)({\AA}) & 84 & 267 & 111 \\
L$_{9000}$ (10$^{36}$erg s$^{-1}${\AA}$^{-1}$) & 1.4 & 0.12 & 0.2
\end{tabular}
%\end{minipage}
\end{table}

\clearpage

%%%%%%%%%%%%%%%%%%%%%%%%%%%%%%%%%%%%%%%%%%%%%%%%%%%%%%%%%%%%%%%%%%%%%%%%%%%%%%%
%
%  REGION CD83 EN NGC 1232 (Z = 0.5 SOLAR + WR FEATURES)
%
%%%%%%%%%%%%%%%%%%%%%%%%%%%%%%%%%%%%%%%%%%%%%%%%%%%%%%%%%%%%%%%%%%%%%%%%%%%%%%%%

\begin{table}
\setcounter{table}{9}
 %\begin{minipage}{60mm}
  \caption{Evolutionary models for region CDT3 in NGC 1232}
  \begin{tabular}{@{}lcccc@{}}
   Parameter & Observations & 2.8 Myr & 4.8 Myr  \\
             &              &         &        \\
 Q(He)/Q(H)  &              &  0.088  & 0.082  \\
 Q(He$^+$)/Q(He)&     & 5.1$\times$10$^{-6}$   & 0.25  \\
 WR/O        &       &  0.0           & 0.027    \\ 
 $<$log U$>$ & -2.72 $\pm$ 0.10 & -2.80 & -2.82  \\
 n$_{e}$ & 223 & 230 & 230 \\  
 12 + log(O/H) & 8.56 $\pm$ 0.14 & 8.58 & 8.55  \\
 log(S/O) & -1.70 $\pm$ 0.07 & -1.70 & -1.71  \\
 log(N/O) & -0.96 $\pm$ 0.04 & -1.00 & -0.96  \\
 3727 [OII]  & 3180 $\pm$ 150 & 4471 & 2868 & \\
 5007 [OIII] & 842 $\pm$ 13 & 823 & 1972  \\
 6584 [NII]  & 892 $\pm$ 30 & 894 & 890  \\
 5755 [NII]  & 8 $\pm$ 2 & 10 & 9  \\
 6717 [SII]  & 333 $\pm$ 10 & 301 & 329  \\
 4072 [SII]  & 48 $\pm$ 6 & 40 & 43 \\
 9069 [SIII] & 229 $\pm$ 10 & 340 & 271 \\
 6312 [SIII] & 8 $\pm$ 2 & 13 & 10  \\
 t(O$^{2+}$) & 0.69 & 0.82 & 0.81  \\
 t(S$^{2+}$) & 0.74 $\pm$ 0.05 & 0.85 & 0.84  \\
 t(S$^{+}$)  & 0.90 $\pm$ 0.06 & 0.88 & 0.84  \\
 t(N$^{+}$)  & 0.86 $\pm$ 0.06 & 0.89 & 0.87  \\
EW(H$\beta$)({\AA}) & 189 & 257 & 60  \\
L$_{9000}$ (10$^{36}$erg s$^{-1}${\AA}$^{-1}$) & 1.6 & 1.9 & 4.9  \\
\end{tabular}
%\end{minipage}
\end{table}

%%%%%%%%%%%%%%%%%%%%%%%%%%%%%%%%%%%%%%%%%%%%%%%%%%%%%%%%%%%%%%%%%%%%%%%%%%%%%%%
%
%      REGION CDT4 EN NGC 1232 (Z = 0.4 SOLAR + WR FEATURES)
%
%%%%%%%%%%%%%%%%%%%%%%%%%%%%%%%%%%%%%%%%%%%%%%%%%%%%%%%%%%%%%%%%%%%%%%%%%%%%%%%

\begin{table}
\setcounter{table}{10}
 %\begin{minipage}{60mm}
  \caption{Evolutionary models for region CDT4 in NGC 1232}
  \begin{tabular}{@{}lcccc@{}}
   Parameter & Observations & 2.8 Myr & 4.8 Myr \\
             &              &         &         \\
 Q(He)/Q(H)  &    &  0.088    & 0.082  \\
 Q(He$^+$)/Q(He)&     & 5.1$\times$10$^{-6}$  & 0.25  \\
 WR/O        &     &   0.0      & 0.027      \\ 
 $<$log U$>$ & -2.72 $\pm$ 0.10 & -2.65 & -2.73 \\
 n$_{e}$ & 118 & 100 & 100 \\  
 12 + log(O/H) & 8.37 $\pm$ 0.12 & 8.45 & 8.45 \\
 log(S/O) & -1.62 $\pm$ 0.07 & -1.60 & -1.65 \\
 log(N/O) & -0.91 $\pm$ 0.06 & -0.95 & -0.95 \\
 3727 [OII]  & 2530 $\pm$ 120 & 3982 & 2582 \\
 5007 [OIII] & 1162 $\pm$ 8 & 1093 & 2193 \\
 6584 [NII]  & 758 $\pm$ 40 & 729 & 730 \\
 5755 [NII]  & 8 $\pm$ 1 & 10 & 9 \\
 6717 [SII]  & 307 $\pm$ 20 & 257 & 305 \\
 9069 [SIII] & 249 $\pm$ 10 & 373 & 283 \\
 6312 [SIII] & 13 $\pm$ 2 & 16 & 12 \\
 t(O$^{2+}$) & 0.84 & 0.86 & 0.85 \\
 t(S$^{2+}$) & 0.87 $\pm$ 0.04 & 0.90 & 0.87 \\
 t(N$^{+}$)  & 0.90 $\pm$ 0.06 & 0.94 & 0.91 \\
EW(H$\beta$)({\AA}) & 138 & 256 & 60 \\
L$_{9000}$ (10$^{36}$erg s$^{-1}${\AA}$^{-1}$) & 7.1 & 2.4 & 6.2
\end{tabular}
%\end{minipage}
\end{table}

\clearpage

%%%%%%%%%%%%%%%%%%%%%%%%%%%%%%%%%%%%%%%%%%%%%%%%%%%%%%%%%%%%%%%%%%%%%%%%%%%%%%%
%
%                 REGION CDT1 EN NGC 1637 (ALTA MET)
%
%%%%%%%%%%%%%%%%%%%%%%%%%%%%%%%%%%%%%%%%%%%%%%%%%%%%%%%%%%%%%%%%%%%%%%%%%%%%%%%

\begin{table}
\setcounter{table}{11}
 %\begin{minipage}{60mm}
  \caption{Evolutionary models for region CDT1 in NGC 1637}
  \begin{tabular}{@{}lcccc@{}}
   Parameter & Observations & 2.8 Myr & 5.9 Myr \\
             &              &         &         \\
 Q(He)/Q(H)  &    &  0.082    & 0.064  \\
 Q(He$^+$)/Q(He)&     & 1.26$\times$10$^{-4}$  & 1.20$\times$10$^{-4}$  \\
 WR/O        &     &  0.24       &    $>$0.9      \\ 
 $<$log U$>$ & -3.30 $\pm$ 0.20 & -3.25 & -3.15 \\
 n$_{e}$ & 100 & 100 & 100 \\  
 12 + log(O/H) & 9.10 & 9.20 & 9.10 \\
 log(S/O) & -1.97 & -1.90 & -1.80 \\
 log(N/O) & -0.88 & -0.87 & -0.87 \\
 3727 [OII]  & 1170 $\pm$ 60 & 1116 & 1031 \\
 5007 [OIII] & 84 $\pm$ 10 & 90 & 90 \\
 6584 [NII]  & 1037 $\pm$ 60 & 1212 & 1039 \\
 6717 [SII]  & 335 $\pm$ 20 & 359 & 379 \\
 9069 [SIII] & 99 $\pm$ 9 & 131 & 141 \\
 t$_{adopt}$  & 0.40 & 0.46 & 0.46 \\
 EW(H$\beta$)({\AA}) & 74 & 137 & 22 \\
L$_{9000}$ (10$^{35}$erg s$^{-1}${\AA}$^{-1}$) & 2.6 & 1.1 & 5.5
\end{tabular}
%\end{minipage}
\end{table}

%%%%%%%%%%%%%%%%%%%%%%%%%%%%%%%%%%%%%%%%%%%%%%%%%%%%%%%%%%%%%%%%%%%%%%%%%%%%%%%
%
%     REGION CDT1 EN NGC 925
%
%%%%%%%%%%%%%%%%%%%%%%%%%%%%%%%%%%%%%%%%%%%%%%%%%%%%%%%%%%%%%%%%%%%%%%%%%%%%%%%

\begin{table}
\setcounter{table}{12}
 %\begin{minipage}{60mm}
  \caption{Evolutionary models for region CDT1 in NGC 925}
  \begin{tabular}{@{}lcccc@{}}
   Parameter & Observations & 5.8 Myr & 5.9 Myr\\
             &              &         &        \\
 Q(He)/Q(H)  &              & 0.124   & 0.076  \\
 Q(He$^+$)/Q(He)&            & 7.58$\times$10$^{-4}$  & 2.86$\times$10$^{-3}$  \\
 WR/O        &              & 0.023   & 0.022  \\   
 $<$log U$>$ & -3.00 $\pm$ 0.30 & -3.15 & -2.95 \\
 n$_{e}$       & $\leq$ 40         & 10    &  10    \\ 
 12 + log(O/H) & 8.52 $\pm$ 0.20 & 8.62 &  8.52  \\
 log(S/O)      & -1.83         & -1.71& -1.71  \\
 log(N/O)      & -1.07         & -1.17& -1.17  \\
 3727 [OII]    & 2880 $\pm$ 110  & 3580 &  3428  \\
 5007 [OIII]   & 849 $\pm$ 14    & 1041 &   809  \\
 6584 [NII]    & 568 $\pm$ 28    & 673  &   577  \\
 6717 [SII]    & 577 $\pm$ 30    & 601  &   390  \\
 9069 [SIII]   & 125 $\pm$ 9     & 233  &   229  \\
 t$_{adopt}$       & 0.86         & 0.74  &   0.78 \\
 EW(H$\beta$)({\AA})    & 22           & 25    &   23   \\
L$_{9000}$ (10$^{35}$erg s$^{-1}${\AA}$^{-1}$) & 9.8 & 2.97 & 3.3  
\end{tabular}
%\end{minipage}
\end{table}

\clearpage

%%%%%%%%%%%%%%%%%%%%%%%%%%%%%%%%%%%%%%%%%%%%%%%%%%%%%%%%%%%%%%%%%%%%%%%%%%%%%%%
%       REGION CDT4 EN NGC 925
%
%%%%%%%%%%%%%%%%%%%%%%%%%%%%%%%%%%%%%%%%%%%%%%%%%%%%%%%%%%%%%%%%%%%%%%%%%%%%%%%

\begin{table*}
\setcounter{table}{13}
 %\begin{minipage}{140mm}
  \caption{Evolutionary models for region CDT4 in NGC 925}
  \begin{tabular}{@{}lcccccc@{}}
   Parameter   & Observations   & 2.8 Myr & 5.9 Myr & Composite\\
               &                &         &         &          \\
 Q(He)/Q(H)    &                & 0.088   & 0.076   &  0.083    \\
 Q(He$^+$)/Q(He)&                & 5.13$\times$10$^{-6}$ & 2.86$\times$10$^{-3}$ &
 1.13$\times$10$^{-3}$\\
 WR/O          &                & 0.00  & 0.022 & \\  
 $<$log U$>$   & -3.00 $\pm$ 0.25 & -3.20 & -3.32 & -2.97\\
 n$_{e}$         & $\leq$ 40         & 10 & 10 & 10\\  
 12 + log(O/H) & 8.41 $\pm$ 0.20  & 8.47  & 8.47 & 8.47\\
 log(S/O)      & -1.83          & -1.71 & -1.71 & -1.71  \\
 log(N/O)      & -1.01          & -1.17 & -1.17  & -1.17 \\
 3727 [OII]    & 2820 $\pm$ 120   & 5133  &  4164 & 4242\\
 5007 [OIII]   & 737 $\pm$ 10     & 257   &   328 & 725\\
 6584 [NII]    & 593 $\pm$ 18     & 764   &   724 & 609\\
 6717 [SII]    & 437 $\pm$ 15     & 508   &   593 & 372\\
 9069 [SIII]   & 160 $\pm$ 10     & 233   &   174 & 248\\
 t$_{adopt}$       & 0.93           & 0.87  &    0.83 & 0.84\\
 EW(H$\beta$)({\AA})   & 47             & 255   &    23 & 48\\
L$_{9000}$ (10$^{35}$erg s$^{-1}${\AA}$^{-1}$) & 4.4 & 0.36 & 1.95 & 2.3 \\
\end{tabular}
%\end{minipage}
\end{table*}
%%%%%%%%%%%%%%%%%%%%%%%%%%%%%%%%%%%%%%%%%%%%%%%%%%%%%%%%%%%%%%%%%%%%%%%%%%%%%%%

%%%%%%%%%%%%%%%%%%%%%%%%%%%%%%%%%%%%%%%%%%%%%%%%%%%%%%%%%%%%%%%%%%%%%%%%%%%%%%%
%       TABLA 15
%
%%%%%%%%%%%%%%%%%%%%%%%%%%%%%%%%%%%%%%%%%%%%%%%%%%%%%%%%%%%%%%%%%%%%%%%%%%%%%%%
\begin{table}
\setcounter{table}{14}
 %\begin{minipage}{60mm}
  \caption{Ionising populations for region CDT1 in NGC 1232}
  \begin{tabular}{@{}lcccc@{}}

 Parameter                 &  2.4 Myr & 7.1 Myr & Composite \\
   
 Q(H) (10$^{50}$ s$^{-1}$)      &  13.8    & 1.38    &   15.2    \\
 Q(He)(10$^{48}$ s$^{-1}$)      &   140    & 0.17    &   140     \\
 L(H$\beta$)(10$^{37}$ erg s$^{-1}$  &   66.1  & 6.61    &    72.7   \\
 M*(10$^{4}$ M$\odot$)           &  2.73    & 15.3    &   18.0    \\
 L(WR)/H$\beta$                &  0.029   &  0.0    &   0.026   \\
 EW(WR)({\AA})                 &  7.3     &  0.0    &   1.1     \\    
 EW(H$\beta$)({\AA})               &  312     &  5.4     &  51  \\
 L(9000)(10$^{36}$ erg s$^{-1}$ {\AA} $^{-1}$)& 0.88 & 4.75 &    5.63  \\   
 $<$log U$>$               & -2.88  & -3.88  & -2.85 \\
 3727 [OII]  & 2119 & 551  & 1904 \\
 4861 H$\beta$   &  1000 & 1000 & 1000 \\
 5007 [OIII] &  294 &   0  &  260 \\
 6584 [NII]  & 1489 & 944  & 1384 \\
 6717 [SII]  &  304 & 474  &  281 \\
 9069 [SIII] &  295 &  15  &  277 \\
 \end{tabular}
%\end{minipage}
\end{table}

\clearpage

%%%%%%%%%%%%%%%%%%%%%%%%%%%%%%%%%%%%%%%%%%%%%%%%%%%%%%%%%%%%%%%%%%%%%%%%%%%%%%%
%       FIGURAS
%
%%%%%%%%%%%%%%%%%%%%%%%%%%%%%%%%%%%%%%%%%%%%%%%%%%%%%%%%%%%%%%%%%%%%%%%%%%%%%%%
%%%%%%%%%%%%%%%%%%%%%%%%%%%%% Figura de la Teq vs age %%%%%%%%%%%%%%%%%%%%%%%%
\begin{figure}
\setcounter{figure}{0}
%\begin{minipage}{80mm}
 \psfig{figure=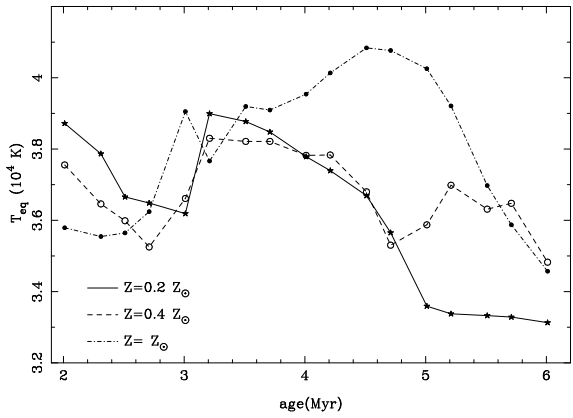,angle=270,width=6.4cm}
\vspace{-2cm}
 \caption{The equivalent temperature, defined as that of the Mihalas
   model atmosphere which provides the same ratio of helium to hydrogen
   ionising photons, {\it versus} age for single ionising clusters of
   different metallicity. }
%\end{minipage}
\end{figure}

%%%%%%%%%%%%%%%%%%%%%%%%%%%%%%%%%%%%%%%%%%%%%%%%%%%%%%%%%%%%%%%%%%%%%%%%%%%%%%%%%%

%%%%%%%%%%%%%%%%%%%%%%%%%%%%% Figura de las DEE de H13 y H3 %%%%%%%%%%%%%%%%%%%%%%%%

\begin{figure}
\setcounter{figure}{1}
 \begin{center}
 \psfig{figure=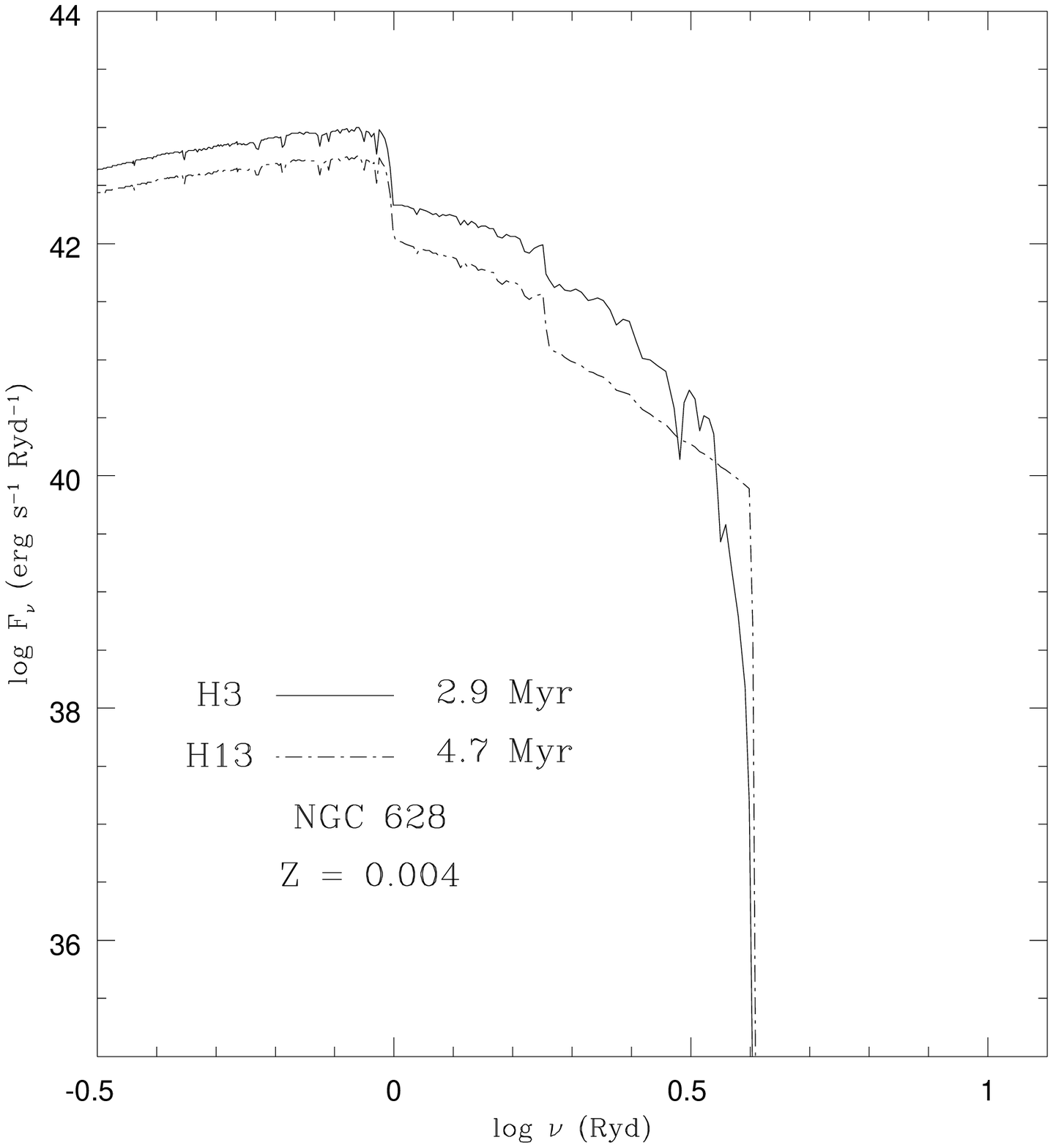,height=8.4cm,width=8.4cm,clip=}
 \caption{Spectral energy distributions of the ionising clusters able
   to reproduce the emission line spectra of regions H13 and H3 in NGC~628}
 \end{center}
\end{figure}

%%%%%%%%%%%%%%%%%%%%%%%%%%%%%%%%%%%%%%%%%%%%%%%%%%%%%%%%%%%%%%%%%%%%%%%%%%%%%%%%%%

\clearpage

%%%%%%%%%%%%%%%%%%%%%%%%%%%%% Figura de las DEE de CDT1 en NGC1232 %%%%%%%%%%%%%%%%%%%

\begin{figure}
\setcounter{figure}{2}
\begin{center}
 \psfig{figure=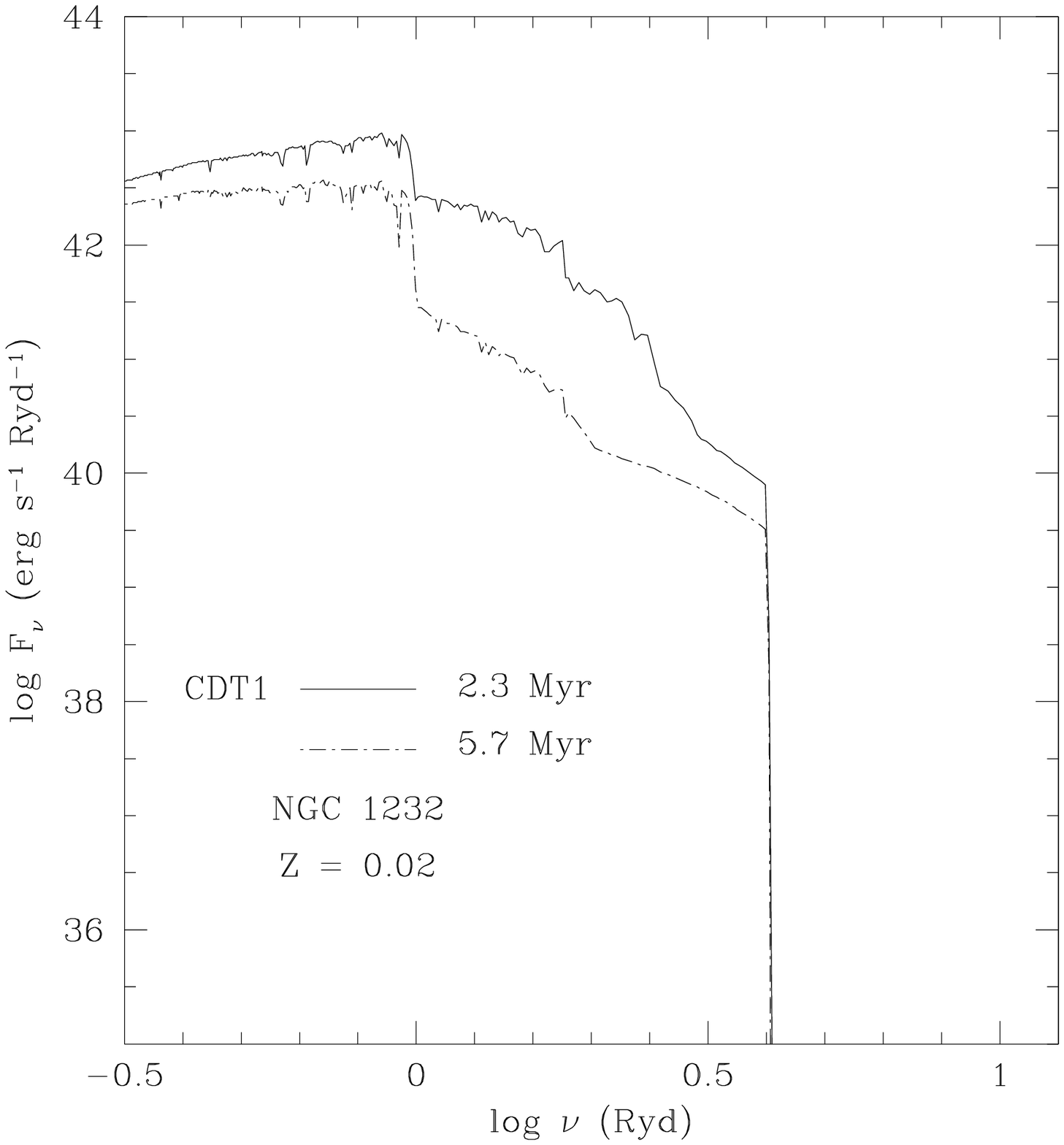,height=8.4cm,width=8.4cm}
 \caption{Spectral energy distributions of the ionising clusters able
   to reproduce the emission line spectrum of region CDT1 in NGC~1232}
\end{center}
\end{figure}

%%%%%%%%%%%%%%%%%%%%%%%%%%%%%%%%%%%%%%%%%%%%%%%%%%%%%%%%%%%%%%%%%%%%%%%%%%%%%%%%%%

%%%%%%%%%%%%%%%%%%%%%%%%%%%%% Figura de las DEE de CDT3 y CDT4 en NGC1232 %%%%%%%%%%%

\begin{figure}
\setcounter{figure}{3}
 \begin{center}
 \psfig{figure=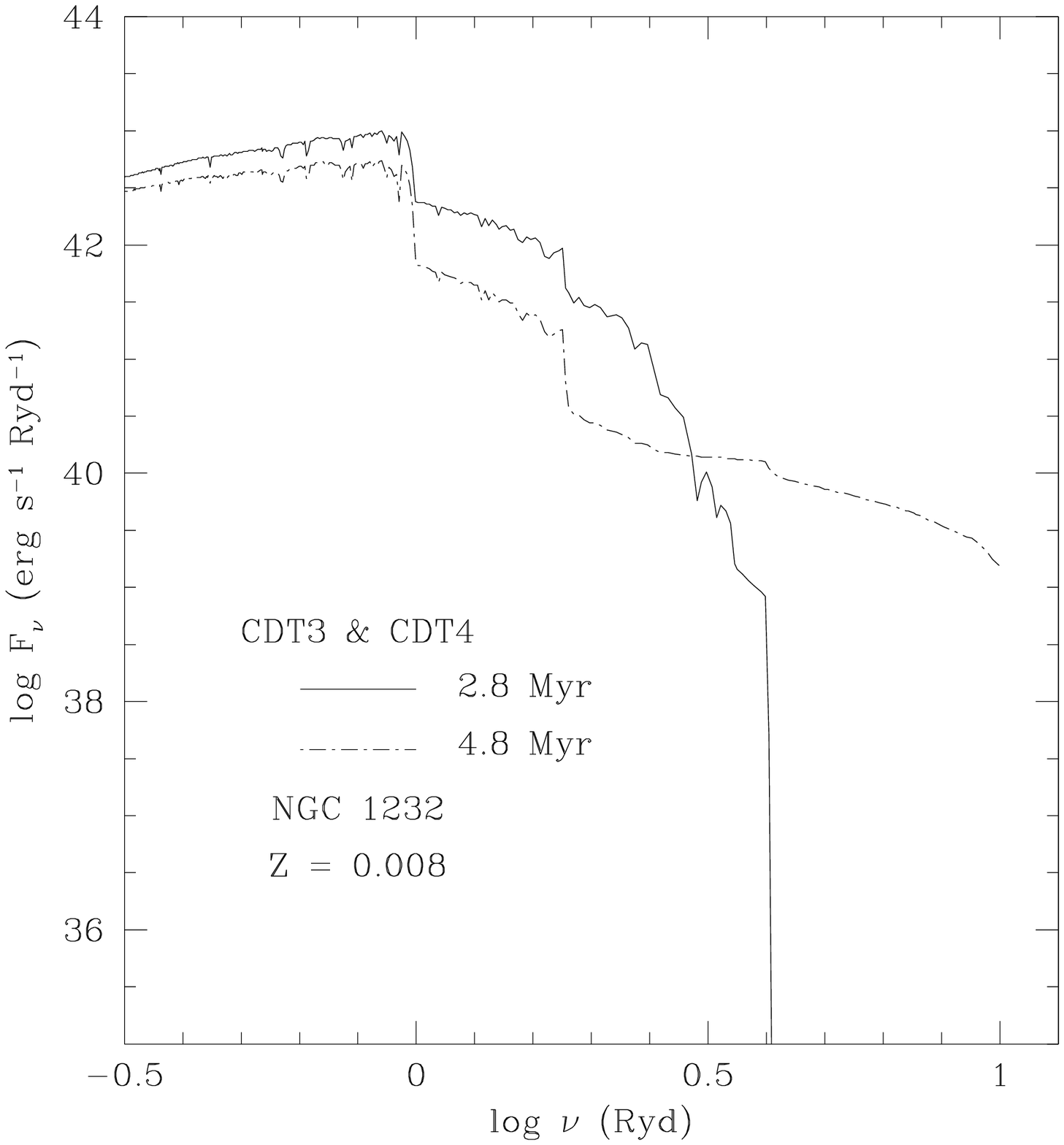,height=8.4cm,width=8.4cm}
 \caption{Spectral energy distributions of the ionising clusters able
   to reproduce the emission line spectra of regions CDT3 and CDT4 in NGC~1232}
 \end{center}
\end{figure}

%%%%%%%%%%%%%%%%%%%%%%%%%%%%%%%%%%%%%%%%%%%%%%%%%%%%%%%%%%%%%%%%%%%%%%%%%%%%%%%%%%

\clearpage

%%%%%%%%%%%%%%%%%%%%%%%%%%%%% Figura de las DEE de CDT3 y CDT4 en NGC1232 %%%%%%%%%%%

\begin{figure}
\setcounter{figure}{4}
 \begin{center}
 \psfig{figure=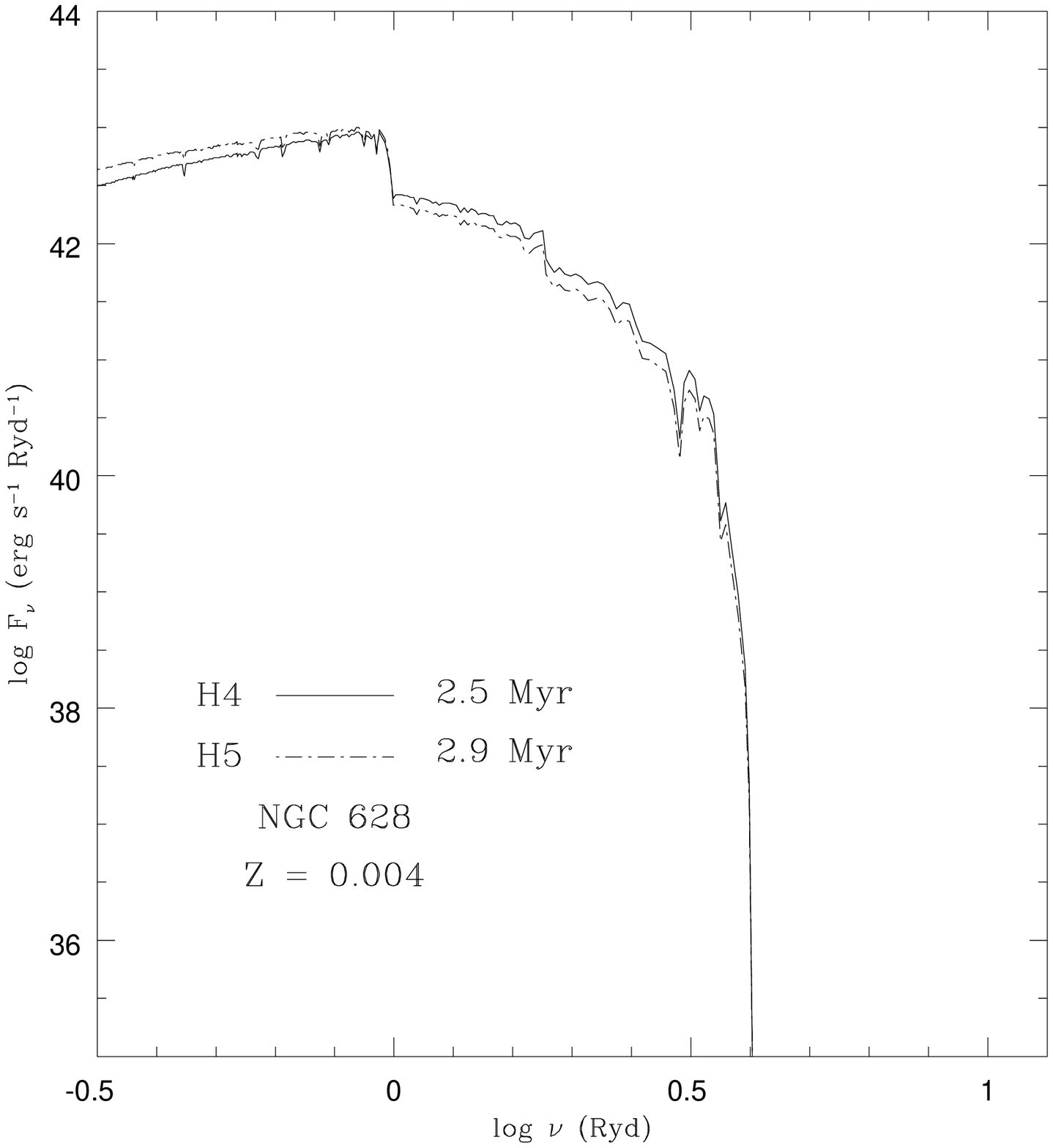,height=8.4cm,width=8.4cm}
 \caption{Spectral energy distributions of the ionising clusters able
   to reproduce the emission line spectra of regions H4 and H5 in NGC~628}
 \end{center}
\end{figure}
%%%%%%%%%%%%%%%%%%%%%%%%%%%%%%%%%%%%%%%%%%%%%%%%%%%%%%%%%%%%%%%%%%%%%%%%%%%%%%%%%%

%%%%%%%%%%%%%%%%%%%%%%%%%%%%% Figura de las DEE de CDT2 en NGC1232 %%%%%%%%%%%%%%%%%%%

\begin{figure}
\setcounter{figure}{5}
 \begin{center}
 \psfig{figure=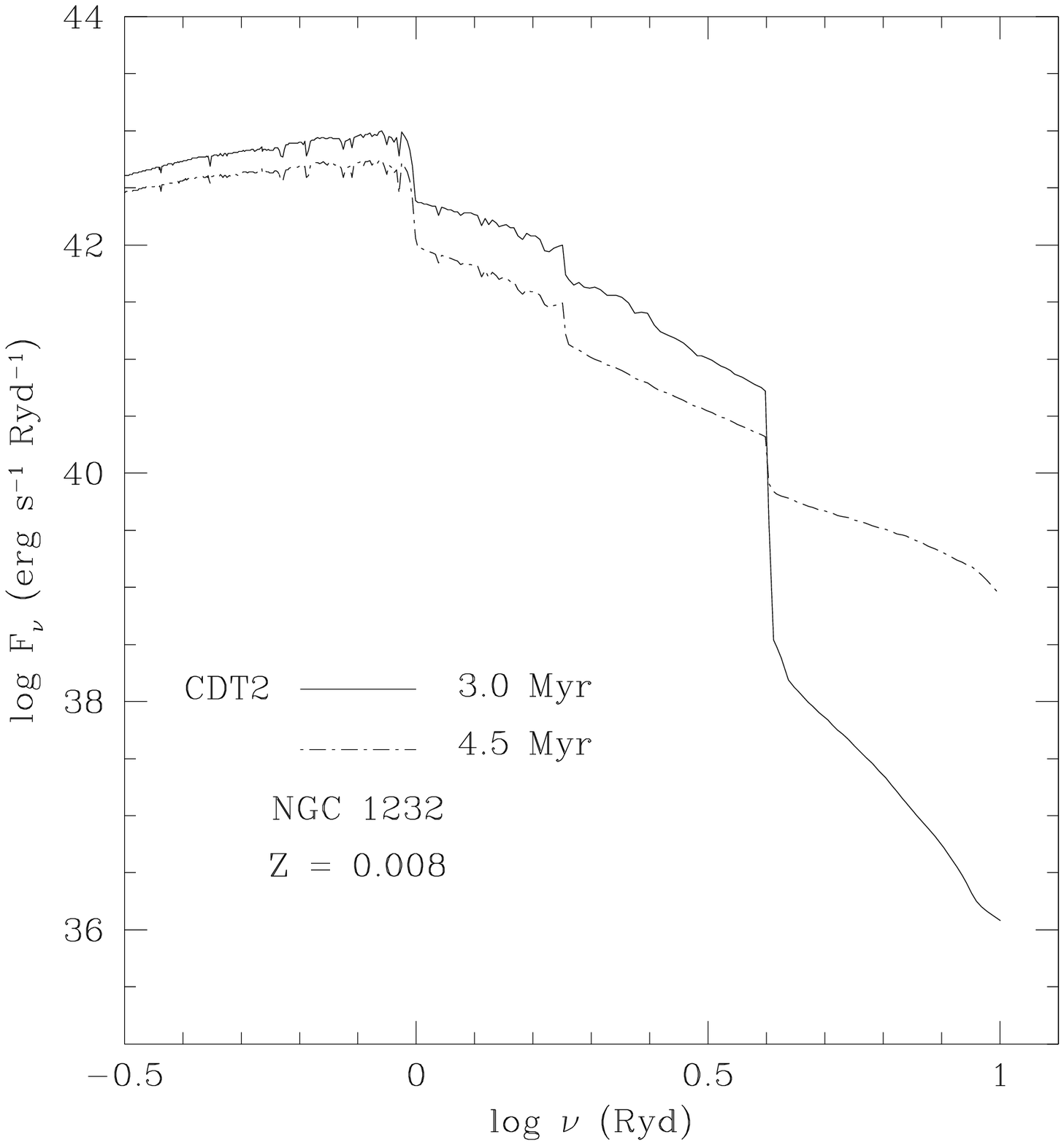,height=8.4cm,width=8.4cm}
 \caption{Spectral energy distributions of the ionising clusters able
   to reproduce the emission line spectrum of region CDT2 in NGC~1232}
 \end{center}
\end{figure}

%%%%%%%%%%%%%%%%%%%%%%%%%%%%%%%%%%%%%%%%%%%%%%%%%%%%%%%%%%%%%%%%%%%%%%%%%%%%%%%%%%

\clearpage

%%%%%%%%%%%%%%%%%%%%%%%%%%%%% Figura de las DEE de CDT1 en NGC1637 %%%%%%%%%%%%%%%%%%%

\begin{figure}
\setcounter{figure}{6}
 \begin{center}
 \psfig{figure=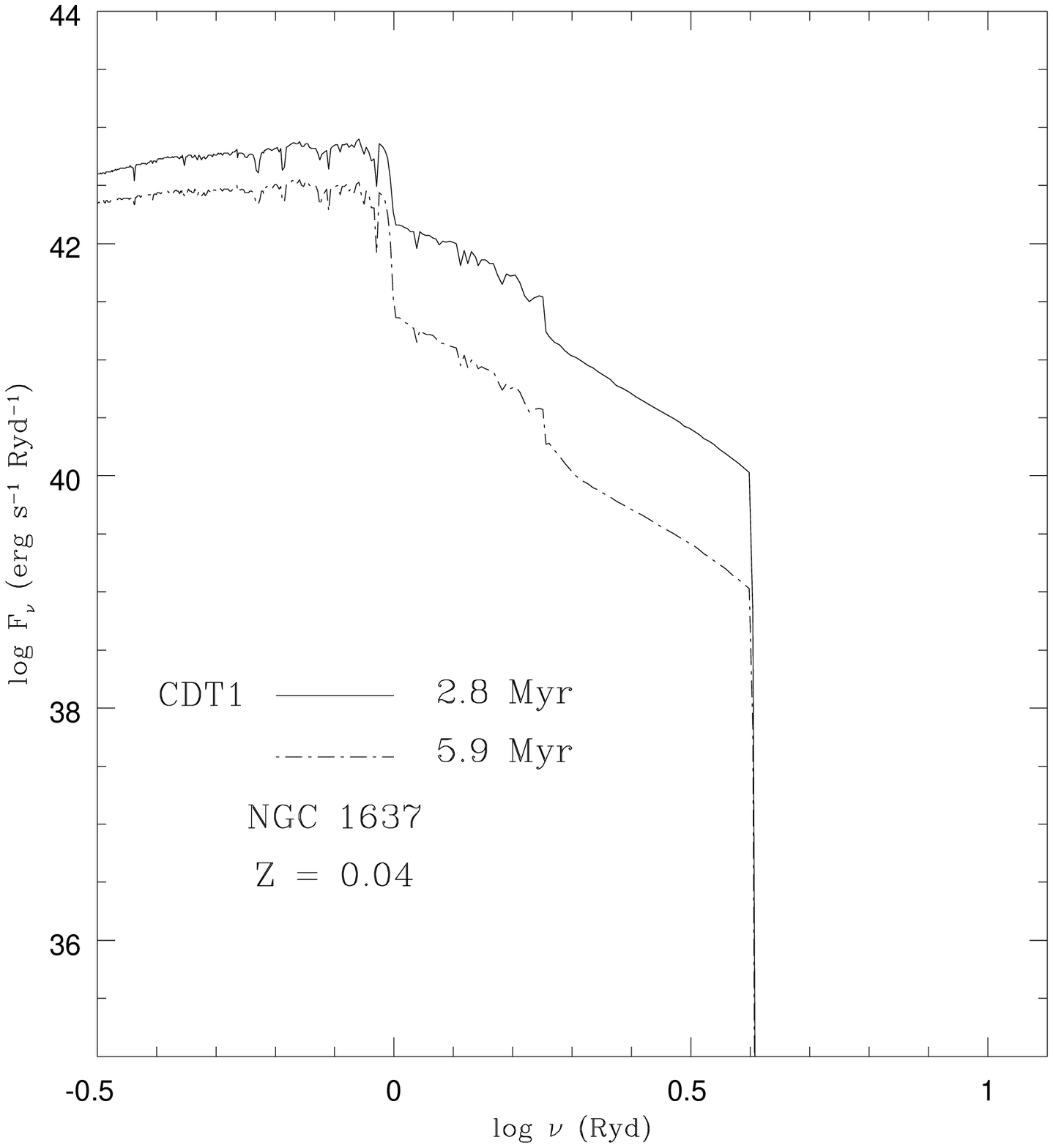,height=8.4cm,width=8.4cm}
 \caption{Spectral energy distributions of the ionising clusters able
   to reproduce the emission line spectrum of region CDT1 in NGC~1637}
 \end{center}
\end{figure}

%%%%%%%%%%%%%%%%%%%%%%%%%%%%%%%%%%%%%%%%%%%%%%%%%%%%%%%%%%%%%%%%%%%%%%%%%%%%%%%%%%

%%%%%%%%%%%%%% Figura 8 con las DEE de CDT1 and CDT4 in NGC925 %%%%%%%%%%%%% 
              
\begin{figure}
\setcounter{figure}{7}
 \begin{center}
 \psfig{figure=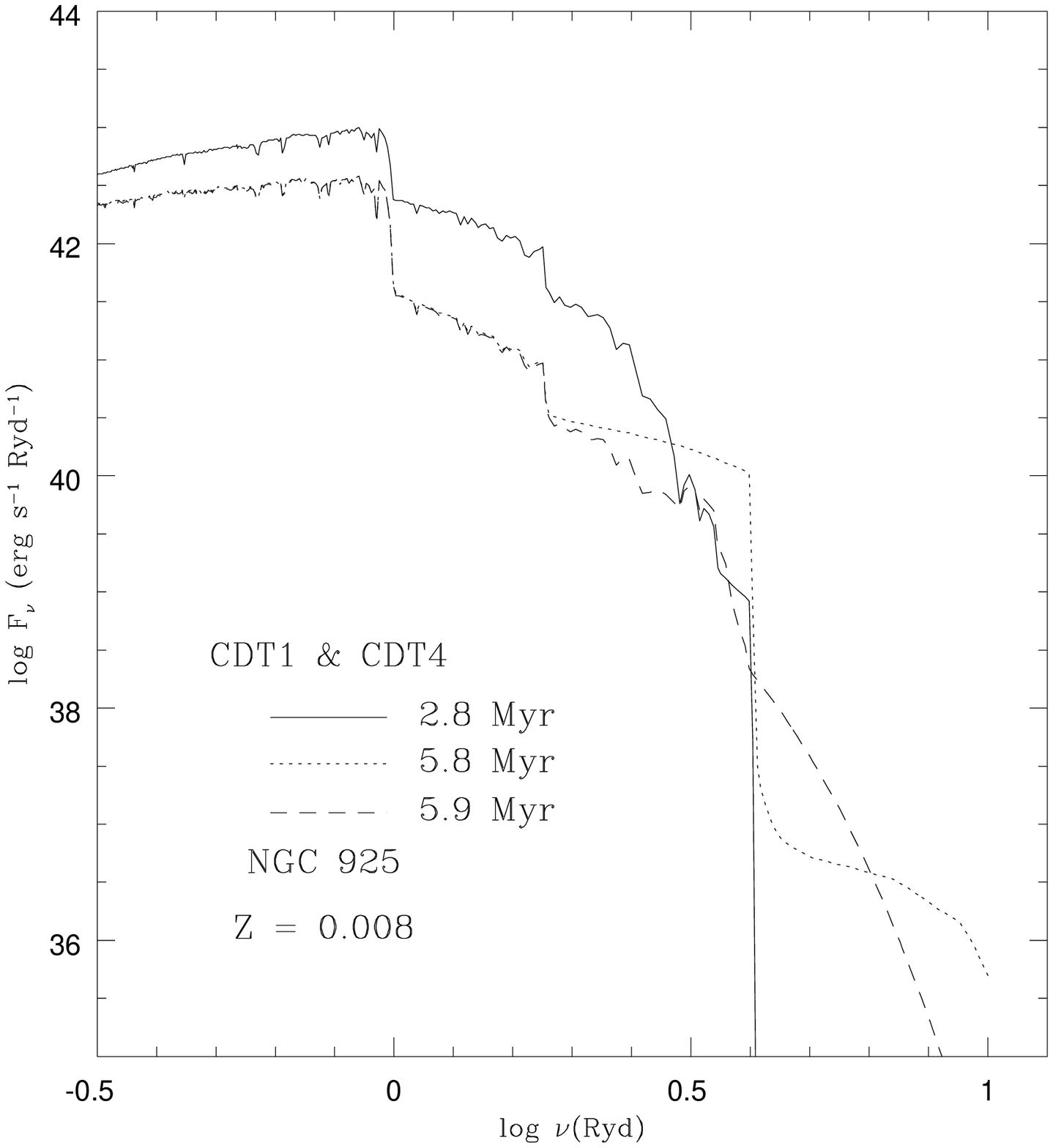,height=8.4cm,width=8.4cm}
 \caption{Spectral energy distributions of the ionising clusters able
   to reproduce the emission line spectra of regions CDT1 and CDT4 in NGC~925}
 \end{center}
\end{figure}

%%%%%%%%%%%%%%%%%%%%%%%%%%%%%%%%%%%%%%%%%%%%%%%%%%%%%%%%%%%%%%%%%%%%%%%%%%%%%%%%%%

\clearpage

%%%%%%%%%%%%%%%%%%%%%%%%%%%%% Figura de las DEE de CDT3 y CDT4 en NGC1232 %%%%%%%%%%%

\begin{figure*}
%\begin{minipage}{160mm}
\setcounter{figure}{8}
 \begin{center}
 \psfig{figure=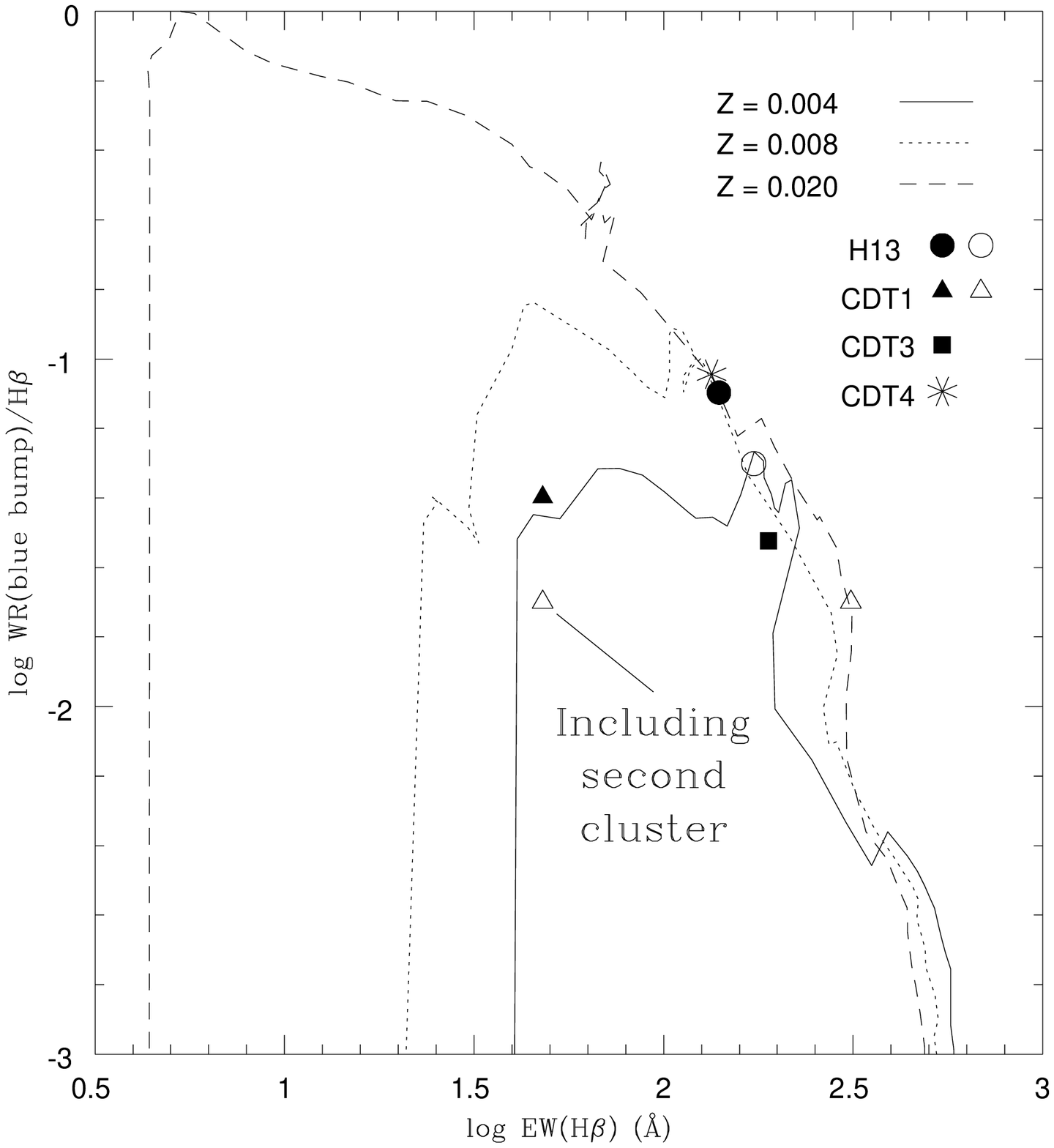,height=8.4cm,width=8.4cm}
\hspace*{0.3cm}
\psfig{figure=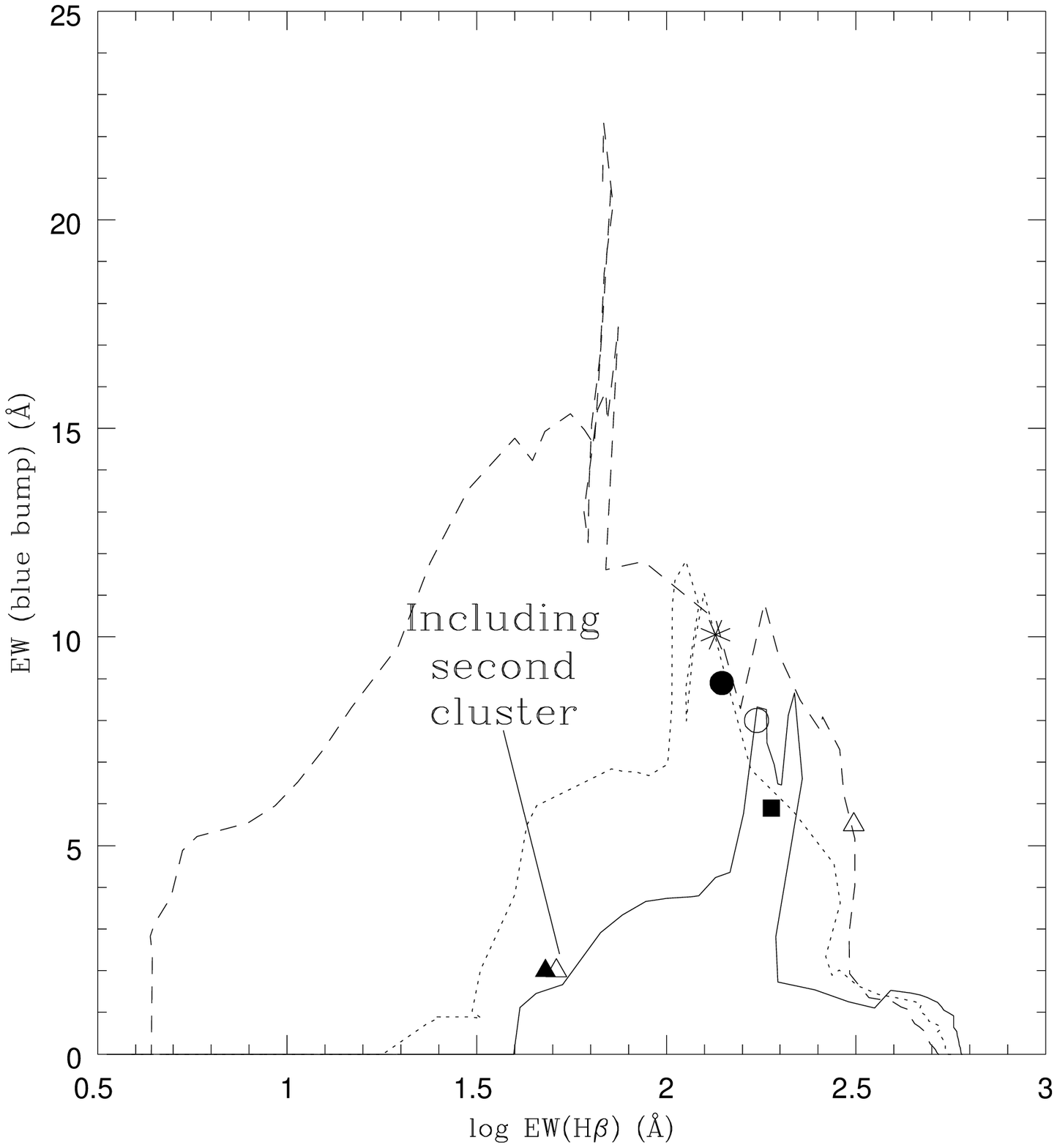,height=8.4cm,width=8.4cm}
 \caption{Relative intensity (left) and equivalent width (right) of the
   WR blue `bump' {\it versus}  H$\beta$ equivalent width for SV98 models of
   three different metallicities as labelled. The data are shown as
   solid and open symbols as explained in the text. }
 \end{center}
%\end{minipage}
\end{figure*}

%%%%%%%%%%%%%%%%%%%%%%%%%%%%%%%%%%%%%%%%%%%%%%%%%%%%%%%%%%%%%%%%%%%%%%%%%%%%%%%%%%
\end{document}